\documentclass[letterpaper,twocolumn,10pt]{article}
\usepackage{usenix-2020-09}

\usepackage{tikz}
\usepackage{amsmath}

\usepackage{xspace}
\usepackage{xstring}
\usepackage{booktabs}
\usepackage{soul}
\usepackage{listings}
\usepackage{subcaption}
\usepackage{xurl}
\usepackage{makecell}
\usepackage{multirow}
\usepackage[framemethod=TikZ]{mdframed}
\usepackage{amssymb}
\usepackage{paralist}
\usepackage{wrapfig}
\usepackage{siunitx}
\usepackage{textpos}

\makeatletter

\renewcommand\paragraph{\@startsection{paragraph}{4}{\z@}%
                                      {1.5ex \@plus 1ex \@minus .2ex}%
                                      {-1em}%
                                      {\normalfont\normalsize\bfseries}}

\renewcommand\subsection{\@startsection{subsection}{2}{\z@}%
	{-2.5ex\@plus -1ex \@minus -.2ex}%
	{1.5ex \@plus .2ex}%
	{\normalfont\normalsize\bfseries}}%

\makeatother

\defaultleftmargin{1.25em}{}{}{}

\usepackage[boxed,noend,noline]{algorithm2e}
\SetAlFnt{\small}
\SetAlCapSkip{0.5em}
\definecolor{mygreen}{rgb}{0,0.6,0}
\definecolor{mygray}{rgb}{0.5,0.5,0.5}
\definecolor{mymauve}{rgb}{0.58,0,0.82}
\definecolor{myblue}{HTML}{02009B}
\definecolor{myred}{HTML}{BF362E}
\definecolor{lightcayenne}{HTML}{B1001C}
\definecolor{lightgrey}{HTML}{EFF0F1}

\makeatletter
\newcommand{\removelatexerror}{\let\@latex@error\@gobble}
\makeatother

\usepackage[firstpage]{draftwatermark}
\SetWatermarkText{\hspace*{6in}\raisebox{7.5in}{\includegraphics{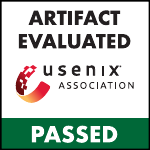}}}
\SetWatermarkAngle{0}

\begin{document}

\date{}

\title{\Large \bf 
Lord of the Ring(s): 
Side Channel Attacks on the \\CPU On-Chip Ring Interconnect Are Practical}

\def\authspace{\hspace{10pt}}

\author{
{\rm Riccardo Paccagnella \authspace Licheng Luo \authspace Christopher W. Fletcher}
\vspace{0.25em}
\\
{University of Illinois at Urbana-Champaign}
} %

\maketitle

\begin{abstract}

We introduce the first microarchitectural side channel attacks that leverage contention on the CPU ring interconnect. %
There are two challenges that make it uniquely difficult to exploit this channel.
First, little is known about the ring interconnect's functioning and architecture.
Second, information that can be learned by an attacker through ring contention is noisy by nature and has coarse spatial granularity. 
To address the first challenge, we perform a thorough reverse engineering of the sophisticated protocols that handle communication on the ring interconnect. 
With this knowledge, we build a cross-core covert channel over the ring 
interconnect with a capacity of over 4 Mbps from a single thread, the largest to date for a cross-core channel not relying on shared memory. 
To address the second challenge, we leverage the fine-grained temporal patterns of ring contention to infer a victim program's secrets.
We demonstrate our attack by extracting key bits from vulnerable EdDSA and RSA implementations, as well as inferring the precise timing of keystrokes typed by a victim user.

\end{abstract}

\ifx\shortversion\undefined
\begin{textblock*}{\textwidth}(88pt,-414pt)
    \textit{This is the extended version of a paper that appears in USENIX Security 2021}
\end{textblock*}
\fi

\section{Introduction}
\label{s:introduction}

Modern computers use multicore CPUs that comprise several heterogeneous, interconnected components often shared across computing units.
While such resource sharing has offered 
significant benefits to efficiency and cost, it has also created an opportunity for new attacks that exploit CPU microarchitectural features. %
One class of these attacks consists of software-based covert channels and side channel attacks.
Through these attacks, an adversary exploits unintended effects (e.g., timing variations) in accessing a particular shared resource to surreptitiously exfiltrate 
data (in the covert channel case) or infer a victim program's secrets
(in the side channel case).
These attacks have been shown to be capable of leaking information 
in numerous contexts.
For example, many cache-based side channel attacks have been demonstrated that can leak sensitive information (e.g., cryptographic keys) in cloud environments~\cite{ristenpart2009hey,zhang2012cross,wu2012whispers,irazoqui2015sa,liu2015last}, web browsers~\cite{oren2015spy,schwarz2017fantastic,gras2017aslr,lipp2017practical} and smartphones~\cite{spreitzer2017systematic,lipp2016armageddon}. 

Fortunately, recent years have also seen an increase in the awareness of such 
attacks, and the availability of countermeasures to mitigate them.
To start with, a large number of existing attacks
(e.g.,~\cite{gras2020absynthe,gras2018translation,aldaya2018port,bhattacharyya2019smotherspectre,percival2005cache,aciicmez2007cheap, evtyushkin2016understanding,evtyushkin2018branchscope}) can be mitigated by disabling simultaneous multi-threading (SMT) and cleansing the CPU microarchitectural state (e.g., the cache) when context switching between different security domains.
Second, cross-core cache-based attacks (e.g.,~\cite{liu2015last,disselkoen2017prime+,gruss2016flush+}) can be blocked by partitioning the last-level cache (e.g., with Intel CAT~\cite{intel-cat-introduction,liu2016catalyst}), and disabling shared memory between processes in different security domains~\cite{vmware-page-sharing}.
The only known attacks that would still work in such a restrictive environment (e.g., DRAMA~\cite{pessl2016drama}) exist outside of the CPU chip.

In this paper, we present the first on-chip, cross-core side channel attack that works despite the above countermeasures.
Our attack leverages contention on the ring interconnect, which is the component that enables communication between the different CPU units (cores, last-level cache, system agent, and graphics unit) on many modern Intel processors.
There are two main reasons that make our attack uniquely challenging.
First, the ring interconnect is a complex architecture with many moving parts.
As we show, understanding how these often-undocumented components interact is an essential prerequisite of a successful attack.
Second, it is difficult to learn sensitive information through the ring interconnect.
Not only is the ring a contention-based channel---requiring precise measurement capabilities to overcome noise---but also it only sees contention due to spatially coarse-grained events such as private cache misses.
Indeed, at the outset of our investigation it was not clear to us whether leaking sensitive information over this channel would even be possible.

To address the first challenge, we perform a thorough reverse engineering of Intel's ``sophisticated ring protocol''~\cite{sandy-bridge-ring-interconnect,lempel2011slides-sandybridge} that handles communication on the ring interconnect.
Our work reveals what physical resources are allocated to what ring agents (cores, last-level cache slices, and system agent) to handle different protocol transactions (loads from the last-level cache and loads from DRAM), and how those physical resources arbitrate between multiple in-flight transaction packets.
Understanding these details is necessary for an attacker to measure victim program behavior.
For example, we find that the ring prioritizes in-flight over new traffic, and that it consists of two independent lanes (each with four physical sub-rings to service different packet types) that service interleaved subsets of agents.
Contrary to our initial hypothesis, this implies that two agents communicating ``in the same direction, on overlapping ring segments'' is not sufficient to
create contention.
Putting our analysis together, we formulate for the first time the necessary and sufficient conditions for two or more processes to contend with each other on the ring interconnect, as well as plausible explanations for what the ring microarchitecture may look like to be consistent with our observations.
We expect the latter to be a useful tool for future work that relies on the CPU uncore.

Next, we investigate the security implications of our findings.
First, leveraging the facts that i) when a process's loads are subject to contention their mean latency is larger than that of regular loads, and ii) an attacker with knowledge of our reverse engineering efforts can set itself up in such a way that its loads are guaranteed to contend with the first process' loads, we build the first cross-core covert channel on the ring interconnect.
Our covert channel does not require shared memory (as, e.g.,~\cite{yarom2014flush,gruss2016flush+}), nor shared access to any uncore structure (e.g., the RNG~\cite{evtyushkin2016covert}).
We show that our covert channel achieves a capacity of up to 4.14 Mbps (518 KBps) from a single thread which, to our knowledge, is faster than all prior channels that do not rely on shared memory (e.g.,~\cite{pessl2016drama}), and within the same order of magnitude as state-of-the-art covert channels that do rely on shared memory (e.g.,~\cite{gruss2016flush+}).

Finally, we show examples of side channel attacks that exploit ring contention.
The first attack extracts key bits~from vulnerable RSA and EdDSA implementations.
Specifically, it abuses mitigations to preemptive scheduling cache attacks to cause the victim's loads to miss in the cache, monitors ring contention while the victim is computing, and employs a standard machine learning classifier to de-noise traces and leak bits.
The second attack targets keystroke timing information (which can be used to infer, e.g., passwords~\cite{zhang2009peeping, song2001timing, kurth2020netcat}).
In particular, we discover that keystroke events cause spikes in ring contention that can be detected by an attacker, even in the presence of background noise.
We show that our attack implementations can leak key bits and keystroke timings with high accuracy. %
We conclude with a discussion of mitigations. %

\begin{figure}[t]
	\centering
	\vspace{2pt}
	\includegraphics[width=\columnwidth]{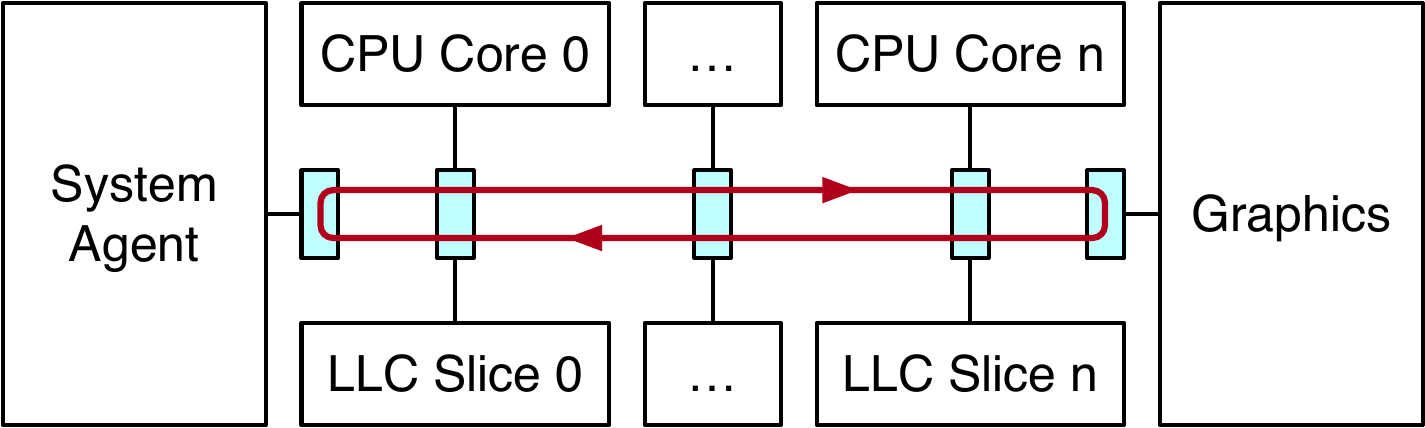}
	\caption{
		Logical block diagram of the ring interconnect on client-class Intel CPUs.
		Ring agents are represented as white boxes, the interconnect is in {\color{lightcayenne} red} and the ring stops are in {\color{cyan} blue}.
		While the positioning of cores and slices on the die varies~\cite{skylake-white-paper}, the ordering of their respective ring stops in the ring is fixed.
	}
	\label{fig:ring-interconnect-diagram}
\end{figure}

\section{Background and Related Work}
\label{s:background}

\paragraph{CPU Cache Architecture}
CPU caches on modern x86 Intel microarchitectures are divided in L1, L2 and L3 (often called last-level cache or LLC).
The L1 and (in most microarchitectures) L2 caches are fast (4 to 12 cycles), small, and local to each CPU core.
They are often referred to as \textit{private caches}.
The LLC is slower (40 to 60 cycles), bigger, and shared across CPU cores.
Since Nehalem-EX~\cite{kottapalli2009nahalem}, the LLC is divided into \textit{LLC slices} of equal size, one per core.

Caches of many Intel CPUs are inclusive, meaning that data contained in the L1 and L2 caches must also reside in the
LLC~\cite{intel-optimization-reference-manual}, and set-associative, meaning that they are divided into a fixed number of \textit{cache sets}, each of which contains a fixed number of \textit{cache ways}.
Each cache way can fit one \textit{cache line} which is typically of 64 bytes in size and represents the basic unit for cache transactions.
The cache sets and the LLC slice in which each cache line is stored are determined by its address bits.
Since the private caches generally have fewer sets than the LLC, it is possible for cache lines that map to different LLC sets to map to the same L2 or L1 set.

When a core performs a load from a memory address, it first looks up if the data associated to that address is available in the L1 and L2. 
If available, the load results in a \textit{hit} in the private caches, and the data is serviced locally.
If not, it results in a \textit{miss} in the private caches, and the core checks if the data is available in the LLC.
In case of an LLC miss, the data needs to be copied from DRAM through the memory controller, which is integrated in the system agent to manage communication between the main memory and the CPU~\cite{intel-optimization-reference-manual}.
Intel also implements hardware prefetching which may result in additional cache lines being copied from memory or from the LLC to the private caches during a single load.

\paragraph{Ring Interconnect}
\label{ss:ring-interconnect}

The \textit{ring interconnect}, often referred to as \textit{ring bus}, is a high-bandwidth on-die interconnect which was introduced by Intel with the Nehalem-EX micro-architecture~\cite{kottapalli2009nahalem} and is used on most Intel CPUs available in the market today~\cite{intel-optimization-reference-manual}.
Shown in Figure \ref{fig:ring-interconnect-diagram}, it is used for intra-processor communication between the cores, the LLC, the system agent (previously known as Uncore) and the GPU.
For example, when a core executes a load and it misses in its private caches (L1-L2), the load has to travel through the ring interconnect to reach the LLC and/or the memory controller.
The different CPU consumers/producers communicating on the ring interconnect are called \textit{ring agents}~\cite{realworldtech-ring}.
Each ring agent communicates with the ring through a \textit{ring stop} (sometimes referred to as \textit{interface block}~\cite{realworldtech-ring}, \textit{node router}~\cite{ausavarungnirun2014design,fallin2011high}, or \textit{ring station}~\cite{saini2013performance}).
Every core shares its ring stop with one LLC slice.
To minimize latency, traffic on the ring interconnect always uses the shortest path, meaning that ring stops can inject/receive traffic in both directions (right or left in our diagram) and always choose the direction with the shortest path to the destination.
Finally, the communication protocol on the the ring interconnect makes use of four physical rings: request, snoop, acknowledge and data ring~\cite{lempel2011slides-sandybridge}.

\subsection{Microarchitectural Side Channels}
\label{ss:side-channel-background}

Most microarchitectural channels can be classified using two criteria.
First, according to the microarchitectural resource that they exploit.
Second, based on the degree of concurrency (also referred to as \textit{leakage channel}) that they rely on~\cite{ge2018survey}.\footnote{Other classifications exist, such as the historical one into storage or timing channels~\cite{dod1985orange}, but our classification is more useful for this paper.}

\paragraph{Target Resource Type}
We distinguish between \textit{eviction-based} (also referred to as \textit{persistent-} or \textit{residual-state}) and \textit{contention-based} (also known as \textit{transient state}) attacks.
Eviction-based channels are stateful: the adversary actively brings the microarchitectural resource into~a~known state, lets the victim execute, and then checks the state of the shared resource again to learn secrets about the victim's execution.
In these attacks, the side effects of the victim's execution leave a footprint that is not undone when the victim code completes.
The root cause of these attacks is the limited storage space of the shared microarchitectural resource.
Examples of shared resources that can be used for eviction-based channels are the L1 data~\cite{percival2005cache,osvik2006cache,lipp2020take} and instruction~\cite{aciiccmez2007yet,zhang2012cross,aciiccmez2008vulnerability} caches, the TLB~\cite{gras2018translation}, the branch target buffer (BTB)~\cite{evtyushkin2016jump, evtyushkin2016understanding} and the last-level cache (LLC)~\cite{liu2015last,yarom2014flush,irazoqui2015sa,gruss2015cache,zhang2014cross, maurice2015c5, gruss2016flush+,kayaalp2016high, disselkoen2017prime+, genkin2017may, schwarz2018keydrown}. 

Contention-based channels are stateless: the adversary passively monitors the latency to access the shared resource and uses variations in this latency to infer secrets about the victim's execution.
In these attacks, the side effects of the victim's execution are only visible while the victim is executing.
The root cause of these attacks is the limited bandwidth capacity of the shared resource.
Examples of resources that can be used for contention-based channels are functional units~\cite{wang2006covert}, execution ports~\cite{aldaya2018port, bhattacharyya2019smotherspectre,gras2020absynthe}, cache banks~\cite{yarom2017cachebleed}, the memory bus~\cite{wu2012whispers} and random number generators~\cite{evtyushkin2016covert}. 
The attack presented in this paper is a contention-based channel.

\paragraph{Leakage Channel}
We further classify attacks as either relying on \textit{preemptive scheduling}, \textit{SMT} or \textit{multicore} techniques.
Preemptive scheduling approaches~\cite{gullasch2011cache,evtyushkin2016understanding, zhang2012cross, neve2006advances, osvik2006cache, ashokkumar2016highly, guanciale2016cache, bruinderink2016flush, evtyushkin2018branchscope, wang2019papp,roy2015design,aciiccmez2007yet}, also referred to as \textit{time-sliced} approaches, consist of the victim and the attacker time-sharing a core.
In these attacks, the victim and the attacker run on the same core and their execution is interleaved.
Simultaneous multithreading (SMT) approaches~\cite{wang2006covert, lipp2020take, aldaya2018port, aciiccmez2007predicting, aciiccmez2007power, osvik2006cache, percival2005cache, gras2018translation} rely on the victim and the attacker executing on the same core in parallel (concurrently).
Multicore approaches~\cite{liu2015last,yarom2014flush,irazoqui2015sa,gruss2015cache,zhang2014cross, maurice2015c5, gruss2016flush+,kayaalp2016high, disselkoen2017prime+, evtyushkin2016covert, sullivan2018microarchitectural, yao2018coherence, yan2019attack} are the most generic with the victim and the attacker running on separate cores.
The attack presented in this paper uses the multicore leakage channel.

\subsection{Side Channel Defenses}

Several defenses to microarchitectural side channels have been proposed.
We focus here on generic approaches.
The most straightforward approach to block a side channel is to disable the sharing of the microarchitectural component on which it relies.
For example, attacks that rely on simultaneous multithreading (SMT) can be thwarted by disabling SMT, which is an increasingly common practice for both cloud providers and end users~\cite{azure-hyperthreading, openbsd-hyperthreading, chromeos-hyperthreading}.
Other approaches propose to partition the shared resource to ensure that its use by the victim cannot be monitored by the attacker~\cite{sprabery2018scheduling, kim2012stealthmem, liu2016catalyst, zhou2016software, wang2014timing}.
For example, Liu et al.~\cite{liu2016catalyst} present a defense to multicore cache attacks that uses Intel CAT~\cite{intel-cat-introduction} to load sensitive victim cache lines in a secure LLC partition where they cannot be evicted by the attacker.
Finally, for channels that rely on preemptive scheduling and SMT, one mitigation approach is to erase the victim's footprint from the microarchitectural state across context switches.
For example, several works proposed to flush the CPU caches on context switches~\cite{sprabery2018scheduling,godfrey2013server, zhang2013duppel, varadarajan2014scheduler, gullasch2011cache, osvik2006cache, braun2015robust, ge2018no, ferraiuolo2018hyperflow, ge2018survey, guanciale2016cache, percival2005cache}.
\section{Reverse Engineering the Ring Interconnect}
\label{s:reverse-engineering}

\begin{figure}[t!]
	\centering
	\includegraphics[width=\columnwidth]{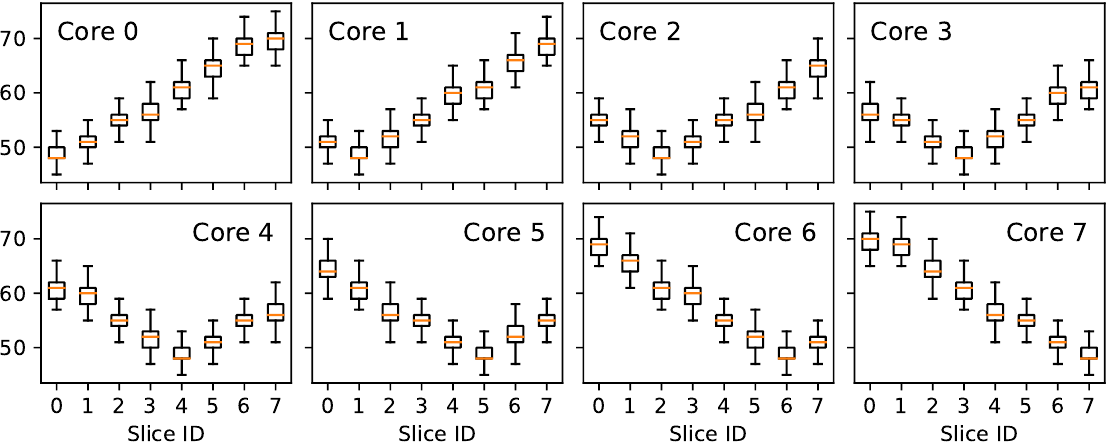}
	\caption{
		Load latency (in cycles) from different LLC slices $s$ (and fixed cache set index $p=5$) on each core $c$ of our Coffee Lake CPU.
		The latency grows when the distance between the core's ring stop and the target LLC slice's ring stop grows.
	}
	\label{fig:latency-coffeelake}
\end{figure}

In this section, we set out to understand the microarchitectural characteristics of the ring interconnect on modern Intel CPUs, with a focus on the necessary and sufficient conditions for an adversary to create and monitor contention on it.
This information will serve as the primitive for our covert channel (Section~\ref{s:covert-channel}) and side channel attacks (Section~\ref{s:side-channel}).

\paragraph{Experimental Setup}

We run our experiments on two machines.
The first one uses an 8-core Intel Core i7-9700 (Coffee Lake) CPU at 3.00GHz.
The second one uses a 4-core Intel Core i7-6700K (Skylake) CPU at 4.00GHz.
Both CPUs have an inclusive, set-associative LLC.
The LLC has 16 ways and 2048 sets per slice on the Skylake CPU and 12 ways and 2048 sets per slice on the Coffee Lake CPU.
Both CPUs have an 8-way L1 with 64 sets and a 4-way L2 with 1024 sets.
We use Ubuntu Server 16.04 with the kernel 4.15 for our experiments.

\subsection{Inferring the Ring Topology}
\label{ss:topology}

\paragraph{Monitoring the Ring Interconnect}
\label{ss:monitoring-program}

We build on prior work~\cite{farshin-slice-aware} and create a monitoring program that measures, from each core, the access time to different LLC slices.
Let $W_{L1}$, $W_{L2}$ and $W_{LLC}$ be the associativities of the L1, L2 and LLC respectively.
Given a core $c$, an LLC slice index $s$ and an LLC cache set index $p$, our program works as follows:
\begin{compactenum}
	\item It pins itself to the given CPU core $c$.
	\item It allocates a buffer of $\ge 400$ MB of memory.\footnote{We found 400 MB to be enough to contain the $W_{LLC}$ addresses of Step~2.}
	\item It iterates through the buffer looking for $W_{LLC}$ addresses that map to the desired slice $s$ and LLC cache set $p$ and stores them into the \textit{monitoring set}.
	      The slice mapping is computed using the approach from Maurice et al.~\cite{maurice2015reverse}, which uses hardware performance counters.
	      This step requires root privileges, but we will discuss later how we can compute the slice mapping also with unprivileged access.
	\item It iterates through the buffer looking for $W_{L1}$ addresses that map to the same L1 and L2 cache sets as the addresses of the monitoring set, but a different LLC cache set (i.e., where the LLC cache set index is not $p$) and stores them into a set which we call the \emph{eviction set}. %
	\item It performs a load of each address of the monitoring set.
	      After this step, all the addresses of the monitoring set should hit in the LLC because their number is equal to $W_{LLC}$.
	      Some of them will hit in the private caches as well.
	\item It evicts the addresses of the monitoring set from the private caches by accessing the addresses of the eviction set.
	      This trick is inspired by previous work~\cite{yan2019attack} and ensures that the addresses of the monitoring set are cached in the LLC, but not in the private caches.
	\item It times loads from the addresses of the monitoring set one at a time using the timestamp counter (\texttt{rdtsc}) and records the measured latencies.
		  These loads will miss in the private caches and hit in the LLC.
		  Thus, they will need to travel through the ring interconnect.
		  Steps 6-7 are repeated as needed to collect the desired number of latency samples.
\end{compactenum}

\paragraph{Results}
We run our monitoring program on each CPU core and collect 100,000 samples of the ``load latency'' from each different LLC slice. %
The results for our Coffee Lake CPU are plotted in Figure \ref{fig:latency-coffeelake}.
\ifx\shortversion\undefined
The results for our Skylake CPU are shown in Appendix~\ref{appendix:reverse-engineering}.
\else
The results for our Skylake CPU are in the extended version~\cite{extendedversion}.
\fi
These results confirm that the logical topology of the ring interconnect on both our CPUs matches the linear topology shown in Figure~\ref{fig:ring-interconnect-diagram}.
That is: %
\begin{mdframed}[backgroundcolor=lightgrey, roundcorner=10pt]
1. The LLC load latency is larger when the load has to travel longer distances on the ring interconnect.
\end{mdframed}

Once this topology and the respective load latencies are known to the attacker, they will be able to map any addresses to their slice by just timing how long it takes to access them and comparing the latency with the results of Figure~\ref{fig:latency-coffeelake}.
As described so far, monitoring latency narrows down the possible slices from $n$ to 2.
To pinpoint the exact slice a line maps to, the attacker can then triangulate from 2 cores.
This does not require root access.
Prior work explores how this knowledge can be used by attackers to reduce the cost of finding eviction sets and by defenders to increase the number of colors in page coloring~\cite{yarom2015mapping,gruss2016flush+}.
What else can an attacker do with this knowledge?
We investigate this question in the next section.

\begin{figure*}[t!]
	\centering
	\includegraphics[width=0.89\linewidth]{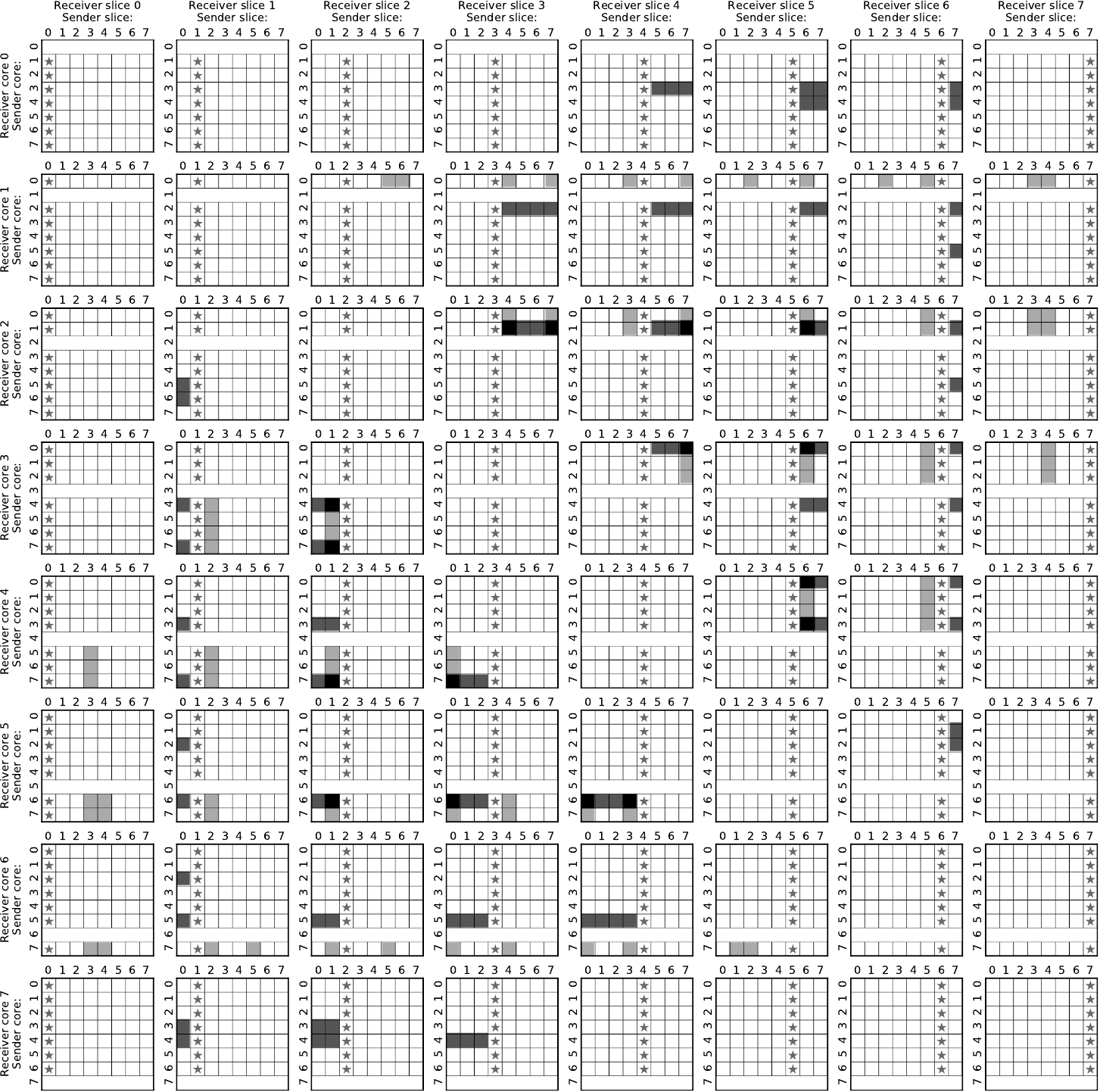}
	\caption{
		Ring interconnect contention heatmap when both the receiver and the sender perform loads that miss in their private caches and hit in the LLC.
		The $y$ axes indicate the core where the sender and the receiver run, and the $x$ axes indicate the target LLC slice from which they perform their loads.
		Cells with a star ($\bigstar$) indicate slice contention (when $R_s = S_s$), 
		while \textbf{gray} cells indicate contention on the ring interconnect (with darker grays indicating larger amounts of contention).
	}
	\vspace{-5pt}
	\label{fig:heatmap-no-miss}
\end{figure*}

\subsection{Understanding Contention on the Ring}

We now set out to answer the question: under what circumstances can two processes contend on the ring interconnect?
To this end, we reverse engineer Intel's ``sophisticated ring protocol''~\cite{sandy-bridge-ring-interconnect,lempel2011slides-sandybridge} that handles communication on the ring interconnect.
We use two processes, a \textit{receiver} and a \textit{sender}.

\paragraph{Measuring Contention}

The \textit{receiver} is an optimized version of the monitoring program described in Section~\ref{ss:topology}, 
that skips Steps 4 and 6 (i.e., does not use an eviction set)
thanks to the following observation: since on our CPUs $W_{LLC} > W_{L1}$ and $W_{LLC} > W_{L2}$, not all the $W_{LLC}$ addresses of the monitoring set can fit in the L1 and L2 at any given time.
For example, on our Skylake machine, $W_{LLC} = 16$ and $W_{L2} = 4$.
Consider the scenario when we access the first 4 addresses of our monitoring set.
These addresses fit in both the private caches and the LLC.
However, we observe that accessing one by one the remaining 12 addresses of the monitoring set evicts the first 4 addresses from the private caches.
Hence, when we load the first addresses again at the next iteration, we still only hit in the LLC.
Using this trick, if we loop through the monitoring set and access its addresses in order, we can always load from the LLC.
To ensure that the addresses are accessed in order, we serialize the loads using pointer chasing, which is a technique also used in prior work~\cite{tromer2010efficient, liu2015last, gras2018translation, wang2019papp}.
Further, to make it less likely to suffer LLC evictions due to external noise, our receiver evenly distributes the $W_{LLC}$ addresses of the monitoring set across two LLC cache sets (within the same slice).
Finally, to amplify the contention signal, our receiver times 4 sequential loads at a time instead of 1.
The bulk of our receiver's code is shown in Listing~\ref{lst:timing_code} (in Appendix~\ref{appendix:extra-data}). %

\paragraph{Creating Contention}
The \textit{sender} is designed to create contention on specific segments on the ring interconnect by ``bombarding'' it with traffic.
This traffic is sent from its core to different CPU components which sit on the ring, such as LLC slices and the system agent.
To target a specific LLC slice, our sender is based on the same code as the receiver.
However, it does not time nor serialize its loads. %
Further, to generate more traffic, it uses a larger monitoring set with $2 \times W_{LLC}$ addresses (evenly distributed across two LLC cache sets).
To target the system agent (\textit{SA}), our sender uses an even larger monitoring set with $N > 2 \times W_{LLC}$ addresses.
Because not all these $N$ addresses will fit in two LLC cache sets, these loads will miss in the cache, causing the sender to communicate with the memory controller (in the SA).

\paragraph{Data Collection}
We use the \textit{receiver} and the \textit{sender} to collect data about ring contention.
For the first set of experiments, we configure the sender to continuously load data from a single LLC slice (without LLC misses).
For the second set of experiments, we configure the sender to always incur misses on its loads from the target LLC slice.
To prevent unintended additional noise, we disable the prefetchers and configure the sender and the receiver to target different cache sets so that they do not interfere through traditional eviction-based attacks.
We refer to the sender's core as $S_c$, the slice it targets as $S_s$, the receiver's core as $R_c$ and the slice it targets as $R_s$.
For both core and slice numbering, we follow the diagram of Figure~\ref{fig:ring-interconnect-diagram}.
For every combination of $S_c$, $S_s$, $R_c$ and $R_s$, we test if running the sender and the receiver concurrently affects the load latency measured by the receiver.
We then compare the results with a baseline, where the sender is disabled.
We say that there is contention when the average load latency measured by the receiver is larger than the baseline.
Figure~\ref{fig:heatmap-no-miss} shows the results of our first experiment, when the sender always hits in the LLC.
Figure~\ref{fig:heatmap-miss} (in Appendix~\ref{appendix:reverse-engineering}) shows the results of our second experiment, when the sender always misses in the LLC.
Both figures refer to our Coffee Lake machine.
The results of our 4-core Skylake machine are a subset of the 8-core Coffee Lake ones (with $R_c < 4 \wedge R_s < 4 \wedge S_c < 4 \wedge S_s < 4$).

\paragraph{Observations When the Sender Hits in the LLC}
First, there is always contention when $S_s = R_s$, irrespective of the sender's and the receiver's positions relative to the LLC slice.
This systematic ``slice contention'' behavior is marked with a $\bigstar$ in Figure~\ref{fig:heatmap-no-miss}, and is likely caused by a combination of i) the sender's and the receiver's loads filling up the slice's request queue (whose existence is mentioned by Intel in~\cite{patra2015fabrics}), thus causing delays to the processing time of the load requests and ii) the sender's and receiver's loads saturating the bandwidth of the shared slice port (which can supply at most 32 B/cycle~\cite{intel-optimization-reference-manual}, or half a cache line per cycle), thus causing delays to the sending of the cache lines back to the cores.
\begin{mdframed}[backgroundcolor=lightgrey, roundcorner=10pt]
	2. When an agent bombards an LLC slice with requests, other agents loading from the same slice observe delays.
\end{mdframed}

Second, when $R_c = R_s$, there is contention iff $S_s = R_s$.
That is, receiver's loads from $R_c = i$ to $R_s = i$ (\emph{core to home slice traffic}) never contend with sender's loads from $S_c \ne i$ to $S_s \ne i$ (\emph{cross-ring traffic}).
This confirms that every core $i$ has a ``home'' slice $i$ that occupies no links on the ring interconnect except for the shared core/slice ring stop~\cite{realworldtech-ring}.
\begin{mdframed}[backgroundcolor=lightgrey, roundcorner=10pt]
	3. A ring stop can service core to home slice traffic and cross-ring traffic simultaneously.
\end{mdframed}

Third, excluding slice contention ($S_s \ne R_s$), there is never contention if the sender and the receiver perform loads in opposite directions. 
For example, there is no contention if the receiver's loads travel from ``left'' to ``right'' ($R_c < R_s$) and the sender's ones from ``right'' to ``left'' ($S_c > S_s$), or vice versa.
The fact that loads in the right/left direction do not contend with loads in the left/right direction confirms that the ring has two physical flows, one for each direction (as per Figure~\ref{fig:ring-interconnect-diagram}).
This observation is supported by Intel in~\cite{lempel2011slides-sandybridge}. 
\begin{mdframed}[backgroundcolor=lightgrey, roundcorner=10pt]
	4. A ring stop can service cross-ring traffic traveling on opposite directions simultaneously.
\end{mdframed}

Fourth, even when the sender and receiver's loads travel in the same direction, there is never contention if the ring interconnect segments between $S_c$ and $S_s$ and between $R_c$ and $R_s$ do not overlap. 
For example, when $R_c = 2$ and $R_s = 5$, there is no contention if $S_c = 0$ and $S_s = 2$ or if $S_c = 5$ and $S_s = 7$.
This is because load traffic on the ring interconnect only travels through the shortest path between the ring stop of the core and the ring stop of the slice.
If the sender's segment does not overlap with the receiver's segment, the receiver will be able to use the full bus bandwidth on its segment.
\begin{mdframed}[backgroundcolor=lightgrey, roundcorner=10pt]
	5. Ring traffic traveling through non-overlapping segments of the ring interconnect does not cause contention.
\end{mdframed}

The above observations narrow us down to the cases when the sender and the receiver perform loads in the same direction and through overlapping segments of the ring.
Before we analyze these cases, recall from Section~\ref{s:background} that the ring interconnect consists of four rings: 1) request, 2) acknowledge, 3) snoop and 4) data rings.
While it is fairly well known that 64~B cache lines are transferred as two packets over the 32~B data ring~\cite{opher-kahn-ring-register,realworldtech-ring, wikichip-sandy-bridge}, little is disclosed about i) what types of packets travel through the other three rings and ii) how packets flow through the four rings during a load transaction.
Intel partially answers (i) in~\cite{patra2015fabrics} and~\cite{intel-xeon-e5-uncore-monitoring} where it explains that the separate rings are respectively used for 1) read/write requests 2) global observation\footnote{Global observations are also known as completion messages~\cite{sundararaman2017cross}.} and response messages, 3) snoops to the cores\footnote{
	Snoops are related to cache coherence. 
	For example, when multiple cores share a cache line in the LLC, the shared LLC can send snoop packets to cores to maintain coherency across copies. Because our sender and receiver do not share any data, their loads should not need to use the snoop ring.} 
and 4) data fills and write-backs.
Further, Intel sheds lights on (ii) in an illustration from~\cite{lempel2011slides-sandybridge} which explains the flow of an LLC hit transaction: 
the transaction starts with a request packet that travels from the core to the target LLC slice\footnote{Cores use a fixed function to map addresses to slices~\cite{wikichip-sandy-bridge,maurice2015reverse,lempel2011slides-sandybridge}.} (\ul{hit flow 1: core$\rightarrow$slice, request});
upon receipt of such packet, the slice retrieves the requested cache line;
finally, it sends back to the core a global observation (GO) message followed by the two data packets of the cache line (\ul{hit flow 2: slice$\rightarrow$core, data and acknowledge}).
\begin{mdframed}[backgroundcolor=lightgrey, roundcorner=10pt]
	6. The ring interconnect is divided into four independent and functionally separated rings.
	A clean LLC load uses the request ring, the acknowledge ring and the data ring.
\end{mdframed}

Importantly, however, our data shows that performing loads in the same direction and sharing a segment of the ring interconnect with the receiver is not a sufficient condition for the sender to create contention on the ring interconnect.

First, the receiver does not see any contention if its traffic envelops the sender's traffic of the ring interconnect (i.e., $R_c < S_c \le S_s < R_s$ or $R_s < S_s \le S_c < R_c$).
For example, when $R_c = 2$ and $R_s = 5$, we see no contention if $S_c = 3$ and $S_s = 4$.
This behavior is due to the distributed arbitration policy on the ring interconnect.
Intel explains it with an analogy, comparing the ring to a train going along with cargo where each ring slot is analogous to a boxcar without cargo~\cite{opher-kahn-ring-register}.
To inject a new packet on the ring, a ring agent needs to wait for a free boxcar.
This policy ensures that traffic on the ring is never blocked but it may delay the injection of new traffic by other agents, because packets already on the ring have priority over new packets.\footnote{This arbitration policy has been previously described also in~\cite{ausavarungnirun2014design, holcomb2014compositional,fallin2011high}.} 
To create contention on a ring, the sender thus needs to inject its traffic into that ring so that it has priority over the receiver's traffic, which can only happen if its packets are injected at stops upstream from the receiver's ones.
\begin{mdframed}[backgroundcolor=lightgrey, roundcorner=10pt]
	7. A ring stop always prioritizes traffic that is already on the ring over new traffic entering from its agents.
	Ring contention occurs when existing on-ring traffic delays the injection of new ring traffic.
\end{mdframed}

Second, even when the sender's traffic is prioritized over the receiver's traffic, the receiver does not always observe contention. 
Let cluster $A = \{0, 3, 4, 7\}$ and $B = \{1, 2, 5, 6\}$.
When the sender has priority on the request ring (on the core$\rightarrow$slice traffic), there is contention if $S_s$ is in the same cluster as $R_s$.
Similarly, when the sender has priority on the data/acknowledge rings (on the slice$\rightarrow$core traffic), there is contention if $S_c$ is in the same cluster as $R_c$.
If the sender has priority on all rings, we observe the union of the above conditions.
This observation suggests that each ring may have two ``lanes'', and that ring stops inject traffic into different lanes depending on the cluster of its destination agent.
As an example for the slice$\rightarrow$core traffic, let $R_c = 2$ ($R_c \in B$) and $R_s = 5$.
In this case, traffic from $R_s$ to $R_c$ travels on the lane corresponding to core cluster B.
When $S_c = 3$ ($S_c \in A$) and $S_s = 7$, traffic from $S_s$ to $S_c$ travels on the lane corresponding to core cluster A.
The two traffic flows thus do not contend.
However, if we change $S_c$ to $S_c = 1$ ($S_c \in B$), the traffic from $S_s$ to $S_c$ also travels on the lane corresponding to core cluster B, thus contending with the receiver.
\begin{mdframed}[backgroundcolor=lightgrey, roundcorner=10pt]
	8. Each ring has two lanes.
	Traffic destined to slices in $A = \{0, 3, 4, 7\}$ travels on one lane, and traffic destined to slices in $B = \{1, 2, 5, 6\}$ travels on the other lane.
	Similarly, traffic destined to cores in $A = \{0, 3, 4, 7\}$ travels on one lane, and traffic destined to cores in $B = \{1, 2, 5, 6\}$ travels on the other lane.
\end{mdframed}

Finally, we observe that the amount of contention that the sender causes when it has priority only on the slice$\rightarrow$core traffic is larger than the amount of contention that it causes when it has priority only on the core$\rightarrow$slice traffic.
This is because i) slice$\rightarrow$core consists of both acknowledge ring and data ring traffic, delaying the receiver on two rings, while core$\rightarrow$slice traffic only delays the receiver on one ring (the request ring) and ii) slice$\rightarrow$core data traffic itself consists of 2 packets per load which occupy more slots (``boxcars'') on its ring, while request traffic likely consists of 1 packet, occupying only one slot on its ring.
Furthermore, the amount of contention is greatest when the sender has priority over both slice$\rightarrow$core and core$\rightarrow$slice traffic.
\begin{mdframed}[backgroundcolor=lightgrey, roundcorner=10pt]
	9. Traffic on the data ring creates more contention than traffic on the request ring. Further, contention is larger when traffic contends on multiple rings simultaneously.
\end{mdframed}

Putting this all together, Figure~\ref{fig:heatmap-no-miss} contains two types of contention: slice contention (cells with a $\bigstar$) and ring interconnect contention (gray cells).
The latter occurs when the sender's request traffic delays the injection of the receiver's request traffic onto the request ring, or the sender's data/GO traffic delays the injection of the receiver's data/GO traffic onto the data/acknowledge rings.
For this to happen, the sender's traffic needs to travel on the same lane, on overlapping segments, and in the same direction as the receiver's traffic, and must be injected upstream from the receiver's traffic. 
Formally, when the sender hits in the LLC cache, contention happens iff:
\begin{equation}
	\small
	\begin{aligned}
		(S_s &= {} R_s) \vee {} \\
		(R_c &< {} R_s) \wedge
		\bigl\{ 
		(S_c < R_c) \wedge (S_s > R_c) \wedge {} \\
		&\bigl[(S_s \in A) \wedge (R_s \in A)
		\vee (S_s \in B) \wedge (R_s \in B) \bigr] \vee {} \\ 
		&(S_s > R_s) \wedge (S_c < R_s) \wedge {} \\ 
		&\bigl[(S_c \in A) \wedge (R_c \in A) \vee (S_c \in B) \wedge (R_c \in B) \bigr] 
		\bigr\} \vee {} \\
		(R_c &> {} R_s) \wedge
		\bigl\{
		(S_c > R_c) \wedge (S_s < R_c) \wedge {} \\
		&\bigl[(S_s \in A) \wedge (R_s \in A)
		\vee (S_s \in B) \wedge (R_s \in B) \bigr] \vee {} \\ 
		&(S_s < R_s) \wedge (S_c > R_s) \wedge {}\\ 
		&\bigl[(S_c \in A) \wedge (R_c \in A) \vee (S_c \in B) \wedge (R_c \in B) \bigr] 
		\bigr\}
	\end{aligned}
	\vspace{6pt}
	\label{eq:ring-contention-no-miss}
\end{equation}

\paragraph{Observations When the Sender Misses in the LLC}
We now report our observations on the results of our second experiment (shown in Figure~\ref{fig:heatmap-miss}), when the sender misses in the LLC.
Note that the receiver's loads still hit in the LLC.

First, we still observe the same slice contention behavior that we observed when the sender hits in the LLC.
This is because, even when the requested cache line is not present in $S_s$, load requests still need to travel from $S_c$ to $S_s$ first~\cite{intel-optimization-reference-manual} and thus still contribute to filling up the LLC slice's request queue creating delays~\cite{patra2015fabrics}.
Additionally, the sender's requests (\ul{miss flow 1: core$\rightarrow$slice, request}) still contend with the receiver's core$\rightarrow$slice request traffic 
when $R_c$, $R_s$, $S_c$ and $S_s$ meet the previous conditions for request ring contention. %
\begin{mdframed}[backgroundcolor=lightgrey, roundcorner=10pt]
	10. Load requests that cannot be satisfied by the LLC still travel through their target LLC slice.
\end{mdframed}

Second, Intel notes that in the event of a cache miss, the LLC slice forwards the request to the system agent (SA) over the same request ring (same request ring \textit{lane} in our terminology)
from which the request arrived~\cite{patra2015fabrics}.
That is, LLC miss transactions include a second request flow from $S_s$ to the SA (\ul{miss flow 2: slice$\rightarrow$SA, request}).
Our data supports the existence of this flow.
We observe contention when the receiver's loads travel from right to left ($R_c > R_s$), $S_s > R_c$, and the sender and the receiver share the respective lane ($R_s$ is in the same cluster as $S_s$).
For example, when $R_c = 5$, $R_s = 2$ ($R_s \in B$) and $S_s = 6$ ($S_s \in B$) the sender's requests from $S_s$ to the SA contend with the receiver's requests from $R_c$ to $R_s$.
One subtle implication of this fact is that the SA behaves differently than the other ring agent types (slices and cores) in that it can receive request traffic on either lane of the request ring.
We find that $S_s$ simply forwards the request (as new traffic) to the SA on the same lane on which it received it from $S_c$, subject to the usual arbitration rules.

We make two additional observations:
i) The amount of contention caused by the slice$\rightarrow$SA flow is smaller than the one caused by the core$\rightarrow$slice flow.
We do not have a hypothesis for why this is the case.
ii) In the special case $S_s = R_c$ ($S_s = 5$ in our example) there is slightly less contention than in the cases where $S_s > R_c$.
This may be because, when asked to inject new traffic by both its core and its slice, the ring stop adopts a round-robin policy rather than prioritizing either party.
Intel uses such a protocol in a recent patent~\cite{intelpatent-roundrobin-mesh}.
\begin{mdframed}[backgroundcolor=lightgrey, roundcorner=10pt]
	11. In case of a miss, an LLC slice forwards the request (as new traffic) to the system agent on the same lane in which it arrived.
	When both a slice and its home core are trying to inject request traffic into the same lane, their ring stop adopts a fair, round-robin arbitration policy.
\end{mdframed}

\setlength{\columnsep}{10pt}%

To our knowledge, no complete information has been disclosed on the subsequent steps of an LLC miss transaction.
We report here our informed hypothesis.
In addition to forwarding the request to the SA, slice $S_s$ also responds to the requesting core $S_c$ with a response packet through the acknowledge ring (\ul{miss flow 3: slice$\rightarrow$core, acknowledge}).
After receiving the request from $S_s$, the SA retrieves the data and sends it to the requesting core $S_c$ preceded by a GO message (\ul{miss flow 4: SA$\rightarrow$core, data and acknowledge}).
The transaction completes when $S_c$ receives the requested data.
\begin{wrapfigure}{r}{0.6\columnwidth}
	\vspace{-2mm}
	\centering
	\includegraphics[width=\linewidth]{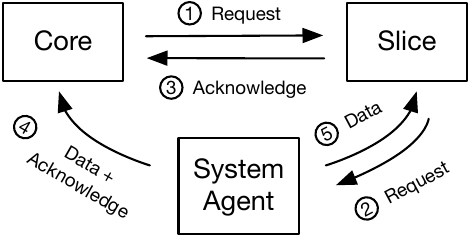}
	\vspace{-6mm}
	\caption{Flows of an LLC miss.}
	\vspace{-4mm}
	\label{fig:miss-flows}
\end{wrapfigure}
To maintain inclusivity, the SA also sends a separate copy of the data to $S_s$ through the data ring (\ul{miss flow 5: SA$\rightarrow$slice, data}).
We summarize the five flows discussed in this part in Figure~\ref{fig:miss-flows}.

The existence of miss flow 4 (SA$\rightarrow$core, data/acknowledge) is supported by the presence of contention when the receiver's loads travel from right to left ($R_c > R_s$), with $S_c > R_s$, and share the respective data/acknowledge ring lanes with the sender. %
For example, there is contention when $R_c = 7$ ($R_c \in A$), $R_s = 2$, $S_c = 3$ ($S_c \in A$) and $S_s = 4$.
Recall from Figure~\ref{fig:ring-interconnect-diagram} that the SA sits on the leftmost ring stop, which implies that traffic injected by the SA always has priority over the receiver's traffic injected by $R_s$.
To corroborate our hypothesis that the SA serves data/acknowledge traffic directly to $S_c$ (and not through $S_s$), we time the load latency of a single LLC miss of the sender with varying $S_c$ and fixed $S_s = 7$.
If our hypothesis held, we would expect a constant latency, because regardless of $S_c$ the transaction would need to travel from $S_c$ to ring stop 7, from ring stop 7 to the SA, and from the SA to $S_c$, which is the same distance regardless of $S_c$; otherwise we would expect to see a decreasing latency as $S_c$ increases.
We measure a fixed latency ($248 \pm 3$ cycles), confirming our hypothesis.
\begin{mdframed}[backgroundcolor=lightgrey, roundcorner=10pt]
	12. The system agent supplies data and global observation messages directly to the core that issued the load.
\end{mdframed}

The existence of miss flow 3 (slice$\rightarrow$core, acknowledge) is supported by the presence of contention in the cases where we previously observed data/acknowledge ring contention with a sender that hits in the LLC.
For example, we observe contention when $R_c = 2$ ($R_c \in B$), $R_s = 6$, $S_c = 5$ ($S_c \in B$) and $S_s = 7$.
However, when the sender misses in the LLC, no data traffic is sent by $S_s$ to $S_c$ (since we saw that data is served to the core directly by the SA).
The contention we observe must then be due to $S_s$ injecting traffic into the acknowledge ring.
Indeed, the amount of contention caused by this acknowledge-only flow is both smaller than the one caused by data/acknowledge flows and equivalent to the one caused by the core$\rightarrow$slice request flow, suggesting that, similarly to the request flow, miss flow 3 may occupy a single slot on its ring.
An Intel patent suggests that miss flow 3 may consist of an ``LLCMiss'' message transmitted by $S_s$ to $S_c$ when the request misses in the LLC~\cite{intelpatent-llcmiss}.
The only remaining question (which we currently cannot answer) is when miss flow 3 occurs: when the miss is detected or when the data is refilled---but both options would cause the same contention. %
\begin{mdframed}[backgroundcolor=lightgrey, roundcorner=10pt]
	13. In the event of a miss, the LLC slice that misses still sends a response packet through the acknowledge ring back to the requesting core.
\end{mdframed}

Finally, the existence of miss flow 5 (SA$\rightarrow$slice, data) is supported by the presence of contention when the receiver's loads travel from right to left ($R_c > R_s$), with $S_s > R_s$, and share the respective lane with the sender.
However, we find a subtle difference in the contention rules of the SA$\rightarrow$slice traffic.
Unlike the SA$\rightarrow$core case, where receiver and sender contend due to traffic of the same type (data and acknowledge) 
being destined to agents of the same type (cores), we now have receiver's and sender's flows of the same type (data) destined to agents of different types (cores and slices, respectively).
In the former case, we saw that the receiver flow and the sender flow share the lane if their destination ring agents are in the same cluster.
In the latter case (which occurs only in this scenario), we observe that the two flows share the lane if their destination ring agents are in different clusters.
This suggests that, as we summarize in Table~\ref{t:lane-agent}, the lanes used to communicate to different clusters may be flipped depending on the destination agent type.
We make two additional observations about miss flow 5.
First, we believe that the SA$\rightarrow$slice traffic only includes data and no acknowledge traffic because the amount of contention that it causes is slightly smaller than the one caused by the SA$\rightarrow$core traffic.
Second, we find that the SA$\rightarrow$slice traffic occurs separately from the SA$\rightarrow$core traffic.
For example, the contention we observe when $R_c = 5$ ($R_c \in B$), $R_s = 2$, $S_c = 4$, $S_s = 3$ ($S_s \in A$) could not occur if the data from the SA had to stop by $S_c$ first.
Also, when the sender contends both on the SA$\rightarrow$slice and SA$\rightarrow$core traffic the contention is larger than the individual contentions, which further supports the independence of the two flows.
\begin{mdframed}[backgroundcolor=lightgrey, roundcorner=10pt]
	14. In the event of a miss, the system agent supplies a separate copy of the data to the missing LLC slice, in order to maintain inclusivity.
	The ring lane used to send data traffic to an LLC slice of one cluster is the same used to send data traffic to a core of the opposite cluster.
\end{mdframed}

\begin{table}[t]
	\centering
	\scriptsize
	\caption{
		Mapping to the ring lane used to send traffic to different agents over any of the four rings.
	}
	\begin{tabular}{l | c c}
		\toprule
		\multirowcell{2}[0ex][l]{Destination\\Ring Agent Type} & \multicolumn{2}{c}{Destination Ring Agent Cluster}                          \\
		                            & {$A = \{0, 3, 4, 7\}$}            & {$B = \{1, 2, 5, 6\}$} \\ \midrule
		Core                        & Lane 1                            & Lane 2                 \\
		LLC Slice                   & Lane 2                            & Lane 1                 \\
		\bottomrule
	\end{tabular}
	\label{t:lane-agent}
\end{table}

To sum up, when the sender misses in the LLC, new ring contention cases occur compared to Equation~\ref{eq:ring-contention-no-miss} due to the extra flows required to handle an LLC miss transaction. %
Formally, contention happens iff:
\begin{equation}
	\small
	\begin{aligned}
		(&S_s = R_s) \vee {} \\
		(&R_c < R_s) \wedge \bigl\{
		(S_c < R_c) \wedge (S_s > R_c) \wedge {} \\
		&\bigl[(S_s \in A) \wedge (R_s \in A)
		\vee (S_s \in B) \wedge (R_s \in B) \bigr] \vee {} \\ 
		&(S_s > R_s) \wedge (S_c < R_s) \wedge {} \\ 
		&\bigl[(S_c \in A) \wedge (R_c \in A) \vee (S_c \in B) \wedge (R_c \in B) \bigr] 
		\bigr\} \vee {} \\
		(&R_c > R_s) \wedge
		\bigl\{
		(S_c > R_c) \wedge (S_s < R_c) \wedge {} \\
		&\bigl[(S_s \in A) \wedge (R_s \in A)
		\vee (S_s \in B) \wedge (R_s \in B) \bigr] \vee {} \\ 
		&(S_s \ge R_c) 
		\wedge \bigl[(S_s \in A) \wedge (R_s \in A)
		\vee (S_s \in B) \wedge (R_s \in B) \bigr] \vee {} \\ 
		&(S_c > R_s) 
		\wedge \bigl[(S_c \in A) \wedge (R_c \in A) \vee (S_c \in B) \wedge (R_c \in B) \bigr] \vee {} \\
		&(S_s > R_s) 
		\wedge \bigl[(S_s \in A) \wedge (R_c \in B) \vee (S_s \in B) \wedge (R_c \in A) \bigr]
		\bigr\}
	\end{aligned}
	\vspace{6pt}
	\label{eq:ring-contention-miss}
\end{equation}

\paragraph{Additional Considerations}

We now provide additional observations on our results.
First, the amount of contention is not proportional to length of the overlapping segment between the sender and the receiver.
This is because, as we saw, contention depends on the presence of full ``boxcars'' passing by the receiver's ring stops when they are trying to inject new traffic, and not on how far away the destination of these boxcars is.

Second, the amount of contention grows when multiple senders contend with the receiver's traffic simultaneously.
This is because multiple senders fill up more slots on the ring, further delaying the receiver's ring stops from injecting their traffic.
For example, when $R_c = 5$ and $R_s = 0$, running one sender with $S_c = 7$ and $S_s = 4$ and one with $S_c = 6$ and $S_s = 3$ creates more contention than running either sender alone.

Third, enabling the hardware prefetchers both amplifies contention in some cases, and causes contention in some new cases (with senders that would not contend with the receiver if the prefetchers were off).
This is because prefetchers cause the LLC or the SA to transfer additional cache lines to the core (possibly mapped to other LLC slices than the one of the requested line), thus filling up more ring slots potentially on multiple lanes. %
Intel itself notes that prefetchers can interfere with normal loads and increase load latency~\cite{vtune-prefetchers-latency}.
We leave formally modeling the additional contention patterns caused by the prefetchers for future work.

Finally, we stress that the contention model we constructed is purely based on our observations and hypotheses from the data we collected on our CPUs.
It is possible that some of the explanations we provided are incorrect.
However, our primary goal is for our model to be useful, and in the next few sections we will demonstrate that it is useful enough to build attacks.

\paragraph{Security Implications}
The results we present bring with them some important takeaways.
First, they suggest an affirmative answer to our question on whether the ring interconnect is susceptible to contention.
Second, they teach us what type of information a receiver process monitoring contention on the ring interconnect can learn about a separate sender process running on the same host.
By pinning itself to different cores and loading from different slices, a receiver may distinguish between the cases when the sender is idle and when it is executing loads that miss in its private caches and are served by a particular LLC slice.
Learning what LLC slice another process is loading from may also reveal some information about the physical address of a load, since the LLC slice an address maps to is a function of its physical address~\cite{wikichip-sandy-bridge,maurice2015reverse,lempel2011slides-sandybridge}.
Further, although we only considered these scenarios, ring contention may be used to distinguish other types on sender behavior, such as communication between the cores and other CPU components (e.g., the graphics unit and the peripherals).
Importantly, however, for any of these tasks the receiver would need to set itself up so that contention with the sender is expected to occur.
Equations~\ref{eq:ring-contention-no-miss} and~\ref{eq:ring-contention-miss} make this possible by revealing the necessary and sufficient conditions under which traffic can contend on the ring interconnect.

\section{Cross-core Covert Channel}
\label{s:covert-channel}

We use the findings of Section~\ref{s:reverse-engineering} to build the first cross-core covert channel to exploit contention on the ring interconnect. 
Our covert channel protocol resembles conventional cache-based covert channels (e.g.,~\cite{liu2015last,yan2019attack}), but in our case the sender and the receiver do not need to share the cache.
The basic idea of the sender is to transmit a bit ``1'' by creating contention on the ring interconnect and a bit ``0'' by idling, thus creating no ring contention.
Simultaneously, the receiver times loads (using the code of Listing~\ref{lst:timing_code}) that travel through a segment of the ring interconnect susceptible to contention due to the sender's loads (this step requires using our results from Section~\ref{s:reverse-engineering}).
Therefore, when the sender is sending a ``1'', the receiver experiences delays in its load latency.
To distinguish a ``0'' from a ``1'' the receiver can then simply use the mean load latency: smaller load latencies are assigned to a ``0'', and larger load latencies are assigned to a ``1''.
To synchronize sender and receiver we use the shared timestamp counter, but our channel could also be extended to use other techniques that do not rely on a common clock (e.g.,~\cite{pessl2016drama, maurice2017hello, wu2012whispers, hunger2015understanding}).

\begin{figure}[t]
	\centering
	\includegraphics[width=1\columnwidth]{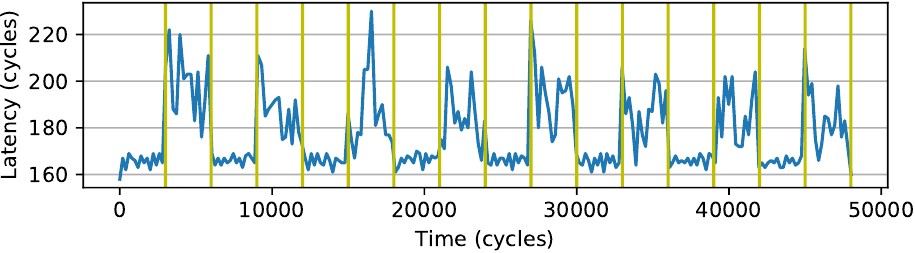}
	\caption{
		Load latency measured by our covert channel receiver when the sender continuously transmits a sequence of zeros (no contention) and ones (contention) on our Coffee Lake machine, with $R_c = 3$, $R_s = 2$, $S_c = 4$, $S_s = 1$ and a transmission interval of 3,000 cycles.
	}
	\label{fig:covert-channel-coffeelake}
\end{figure}

To make the covert channel fast, we leverage insights from Section~\ref{s:reverse-engineering}.
First, we configure the receiver to use a short segment of the ring interconnect.
This allows the receiver to issue more loads per unit time due to the smaller load latency, without affecting the sender's ability to create contention.
Second, we set up the sender to hit in the LLC and use a configuration of $S_c$ and $S_s$ where, based on Equation~\ref{eq:ring-contention-no-miss}, it is guaranteed to contend with the receiver both on its core$\rightarrow$slice traffic and on its slice$\rightarrow$core one.
Contending on both flows allows the sender to amplify the difference between a 0 (no contention) and a 1 (contention).
Third, we leave the prefetchers on, as we saw that they enable the sender to create more contention.

\begin{figure}[t]
	\centering
	\includegraphics[width=1\columnwidth]{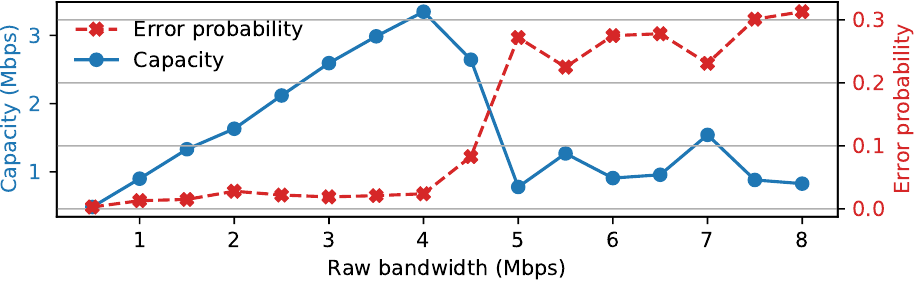}
	\caption{
		Performance of our covert channel implementation on Coffee Lake, reported using raw bandwidth (bits transmitted per second), error probability (percentage of bits received wrong), and channel capacity, which takes into account both bandwidth and error probability to evaluate performance under the binary symmetric channel model (as in, e.g.,~\cite{pessl2016drama,okhravi2010design}).
	}
	\label{fig:covert-channel-capacity-coffeelake}
\end{figure}

We create a proof-of-concept implementation of our covert channel, where the sender and the receiver are single-threaded and agree on a fixed bit transmission interval.
Figure~\ref{fig:covert-channel-coffeelake} shows the load latency measured by the receiver on our Coffee Lake 3.00 GHz CPU, given receiver and sender configurations $R_c = 3$, $R_s = 2$ and $S_c = 4$, $S_s = 1$, respectively.
For this experiment, the sender transmits a sequence of alternating ones and zeros with a transmission interval of 3,000 cycles (equivalent to a raw bandwidth of 1 Mbps).
The results show that ones (hills) and zeros (valleys) are clearly distinguishable.
To evaluate the performance and robustness of our implementation with varying transmission intervals, we use the channel capacity metric (as in~\cite{pessl2016drama,okhravi2010design}).
This metric is computed by multiplying the raw bandwidth with $1 - H(e)$, where $e$ is the probability of a bit error and $H$ is the binary entropy function.
Figure~\ref{fig:covert-channel-capacity-coffeelake} shows the results on our Coffee Lake CPU, with a channel capacity that peaks at 3.35 Mbps (418 KBps) given a transmission interval of 750 cycles (equivalent to a raw bandwidth of 4 Mbps).
To our knowledge, this is the largest covert channel capacity of all existing cross-core covert channels that do not rely on shared memory to date (e.g.,~\cite{pessl2016drama, wu2012whispers}).
We achieve an even higher capacity of 4.14 Mbps (518 KBps) on our Skylake 4.00 GHz CPU by using a transmission interval of 727 cycles, and show the results in Appendix~\ref{appendix:covert-channel}.

Finally, we remark that while our numbers represent a real, reproducible end-to-end capacity, they were collected in the absence of background noise.
Noisy environments may reduce the covert channel performance and require including in the transmission additional error correction codes (as in, e.g.,~\cite{evtyushkin2016covert,maurice2017hello,gruss2016flush+}), that we do not take into account.

\section{Cross-core Side Channels}
\label{s:side-channel}

In this section, we present two examples of side channel attacks that exploit contention on the ring interconnect.

\paragraph{Basic Idea}
In both our attacks, we implement the attacker using the technique described in Section~\ref{s:reverse-engineering} (cf. Listing~\ref{lst:timing_code}).
The attacker (receiver) issues loads that travel over a fixed segment of the ring interconnect and measures their latency. %
We will refer to each measured load latency as a \textit{sample}, and to a collection of many samples (i.e., one run of the attacker/receiver)
as a \textit{trace}.
If during an attack the victim (sender) performs memory accesses that satisfy the conditions of Equations~\ref{eq:ring-contention-no-miss} and \ref{eq:ring-contention-miss} to contend with the attacker's loads, the attacker will measure longer load latencies.
Generally, the slices accessed by an unwitting victim will be uniformly distributed across the LLC~\cite{intel-optimization-reference-manual}.
Therefore, it is likely that some of the victim's accesses will contend with the attacker's loads. %
If the delays measured by the attacker can be attributed to a victim's secret, the attacker can use them as a side channel.

\paragraph{Threat Model and Assumptions}
\label{ss:threat-model}
We assume that SMT is off~\cite{azure-hyperthreading, openbsd-hyperthreading, chromeos-hyperthreading} and that multicore cache-based attacks are not possible (e.g., due to partitioning the LLC~\cite{intel-cat-introduction, liu2016catalyst, sprabery2018scheduling} and disabling shared memory across security domains~\cite{zhou2016software, vmware-page-sharing}).
For our attack on cryptographic code, we also assume that i) the administrator has configured the system to cleanse the victim's cache footprint on context switches (to block cache-based preemptive scheduling attacks~\cite{sprabery2018scheduling,godfrey2013server, zhang2013duppel, varadarajan2014scheduler, ge2018survey, gullasch2011cache, braun2015robust, osvik2006cache, guanciale2016cache, ge2018no, ferraiuolo2018hyperflow,percival2005cache}) and ii) the attacker can observe multiple runs of the victim.
We assume an attacker who has knowledge of the contention model (Section~\ref{s:reverse-engineering}) for the victim's machine and can run unprivileged code on the victim's machine itself. %

\subsection{Side Channel Attack On Cryptographic Code}
\label{s:side-channel-crypto}

\begin{wrapfigure}{R}{0.45\columnwidth}
	\removelatexerror
	\vspace{-3mm}
	\centering
	\begin{algorithm}[H]

		\ForEach{\textnormal{bit} $b$ \textnormal{\textbf{in}} \textnormal{key} $k$}{
			\textit{E1}()\;
			\If{$b == 1$} {
				\textit{E2}()\;
			}
		}
		
		\vspace{-4pt}
		\parbox{.9\linewidth}{\caption{Key-dependent control~flow.}}
		\label{a:simplified-victim}
	\end{algorithm}
	\vspace{-2mm}
\end{wrapfigure}
Our first attack targets a victim that follows the pseudocode of Algorithm~\ref{a:simplified-victim}, where \textit{E1} and \textit{E2} are separate functions executing different operations on some user input (e.g., a ciphertext).
This is a common pattern in efficient implementations of cryptographic primitives that is exploited in many existing side channel attacks against, e.g., RSA~\cite{percival2005cache, yarom2014flush, gras2020absynthe, yan2019attack}, ElGamal~\cite{liu2015last, zhang2012cross}, DSA~\cite{pereida2016make}, ECDSA~\cite{benger2014ooh} and EdDSA~\cite{gras2018translation, gras2020absynthe}.

Let us consider the first iteration of the victim's loop, and, for now, assume that the victim starts from a cold cache, meaning that its code and data are uncached (no prior executions).
When the victim executes \textit{E1} for the first time, it has to load code and data words used by \textit{E1} into its private caches, through the ring interconnect.
Then, there are 2 cases: when the first key bit is 0 and when it is 1.
When the first bit is $0$, the victim's code skips the call to \textit{E2} after \textit{E1} and jumps to the next loop iteration by calling \textit{E1} again.
At this second \textit{E1} call, the words of \textit{E1} are already in the private caches of the victim, since they were just accessed.
Therefore, the victim does not send traffic onto the ring interconnect during the second call to \textit{E1}.
In contrast, when the first bit is $1$, the victim's code calls \textit{E2} immediately after the first \textit{E1}.
When \textit{E2} is called for the first time, its code and data words miss in the cache and loading them needs to use the ring interconnect.
The attacker can then infer whether the first bit was $0$ or $1$ by detecting whether \textit{E2} executed after \textit{E1}.
Contention peaks following \textit{E1}'s execution imply that \textit{E2} executed and that the first secret bit was $1$, while no contention peaks following \textit{E1}'s execution imply that the call to \textit{E1} was followed by another call to \textit{E1} and that the first secret bit was $0$.

We can generalize this approach to leaking multiple key bits by having the attacker interrupt/resume the victim using preemptive scheduling techniques~\cite{gullasch2011cache, zhang2012cross, neve2006advances, osvik2006cache, ashokkumar2016highly, guanciale2016cache, bruinderink2016flush, evtyushkin2018branchscope, wang2019papp, evtyushkin2016understanding, aciiccmez2007yet, roy2015design}.
Let $T_\textit{E1}$ be the median time that the victim takes to execute \textit{E1} starting from a cold cache and $T_\textit{E1+E2}$ be the median time that the victim takes to execute \textit{E1} followed by \textit{E2} starting from a cold cache.
The complete attack works as follows: the attacker starts the victim and lets it run for $T_\textit{E1+E2}$ cycles while concurrently monitoring the ring interconnect.
After $T_\textit{E1+E2}$ cycles, the attacker interrupts the victim and analyzes the collected trace to infer the first secret bit with the technique described above.
Interrupting the victim causes a context switch during which the victim’s cache is cleansed before yielding control to the attacker (cf. Threat Model). 
As a side effect, this brings the victim back to a cold cache state.
If the trace reveals that the first secret bit was 1, the attacker resumes the victim (that is now at the beginning of the second iteration) and lets it run for $T_\textit{E1+E2}$ more cycles, repeating the above procedure to leak the second bit.
If the trace reveals that the first secret bit was 0, the attacker stops the victim (or it lets it finish the current run), starts it again from the beginning, lets it run for $T_\textit{E1}$ cycles, and then interrupts it.
The victim will now be at the beginning of the second iteration, and the attacker can repeat the above procedure to leak the second bit.
The attacker repeats this operation until all the key bits are leaked.
In the worst case, if all the key bits are zeros, our attack requires as many runs of the victim as the number of bits of the key.
In the best case, if all the key bits are ones, it requires only one run of the victim.

\paragraph{Implementation}
We implement a proof-of-concept (POC) of our attack against RSA and EdDSA.
Like prior work~\cite{guanciale2016cache, bruinderink2016flush, evtyushkin2018branchscope,wang2019papp, evtyushkin2016understanding, aciiccmez2007yet, aciiccmez2008vulnerability}, our POC simulates the preemptive~scheduling attack by allowing the attacker to be synchronized with the target iteration of the victim's loop.\footnote{
	Practical implementations of preemptive scheduling techniques (e.g.,~\cite{gullasch2011cache,roy2015design,ashokkumar2016highly,neve2006advances}) are orthogonal to this paper and discussed in Section~\ref{s:discussion}.
}
Further, our POC simulates cache cleansing
by flushing the victim's memory before executing the target iteration.
It does this by calling \texttt{clflush} on each cache line 
making up
the victim's mapped pages (available in \texttt{/proc/[pid]/maps}).\footnote{
	\ifx\shortversion\undefined
	We consider other cache cleansing approaches in Appendix~\ref{appendix:side-channel}, and discuss the implications of this requirement in Section~\ref{s:discussion}.
	\else
	We consider other cache cleansing approaches in the extended version~\cite{extendedversion}, and discuss the implications of this requirement in Section~\ref{s:discussion}.
	\fi
}
Our POC considers the worst-case scenario described above and leaks one key bit per run of the victim.
To simplify the process of inferring a key bit from each resulting trace, our POC uses a Support Vector Machine classifier (SVC).
Note that while the RSA and EdDSA implementations we consider are already known to be vulnerable to side channels, we are the first to show that they leak over the ring interconnect channel specifically.

\begin{figure}[t]
	\begin{subfigure}{\columnwidth}
		\centering
		\includegraphics[width=.92\columnwidth]{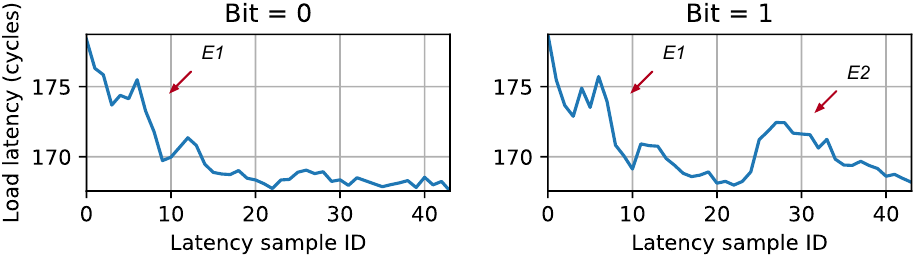}
		\caption{
			Results for the RSA victim.
			When $bit=1$, the attacker sees an additional contention peak between samples 20 and 40.\protect\footnotemark{}
		}
		\label{fig:side-channel-rsa-dreadnought}
	\end{subfigure}
	
	\vspace{10pt}

	\begin{subfigure}{\columnwidth}
		\centering
		\includegraphics[width=.92\columnwidth]{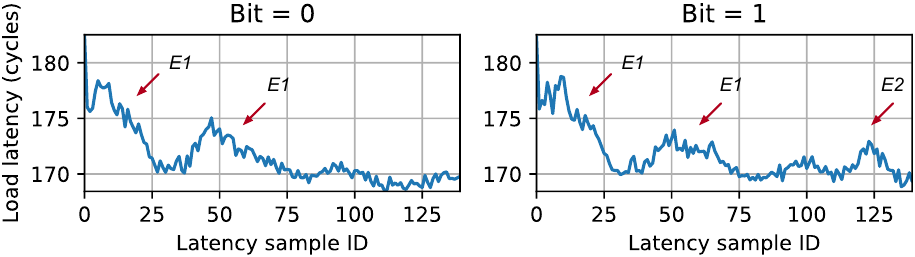}
		\caption{
			Results for the EdDSA victim.
			When $bit = 1$, the attacker sees an additional peak after the 100-th sample.\footnotemark{}
		}
		\label{fig:side-channel-eddsa-dreadnought}
	\end{subfigure}

	\caption{
		Latencies measured by the attacker during a victim's iteration, with $R_c = 2$, $R_s = 1$, and $S_c = 5$ (on Coffee Lake).
	}
	\label{fig:dreadnought-side-channels}
\end{figure}

\paragraph{Results for RSA}
We target the RSA decryption code of libgcrypt 1.5.2 which uses the secret-dependent square-and-multiply method in its modular exponentiation function \texttt{\_gcry\_mpi\_powm}. 
This pattern matches the one of Algorithm~\ref{a:simplified-victim}, with \textit{E1} representing the squaring phase, executed unconditionally, and \textit{E2} representing the multiplication phase, executed conditionally only on 1-valued bits of the key.

We configure the attacker (receiver) on core $R_c = 2$, timing loads from $R_s = 1$, and experiment with different victim (sender) cores $S_c$.
Figure~\ref{fig:side-channel-rsa-dreadnought} shows traces collected by the attacker to leak one key bit of the victim, when $S_c = 5$. %
To better visualize the difference between a 0 bit and a 1 bit, the traces are averaged over 100 runs of the victim.\footnote{Note that, however, our classifier uses a single raw trace as input.}
As expected, we observe that both traces start with peaks, corresponding to the first call to \textit{E1} loading its code and data words from the memory controller through the ring interconnect.
However, only when the secret bit is 1 do we observe an additional peak on the right-hand side of the plot.
This additional peak corresponds to the call to \textit{E2}.
\ifx\shortversion\undefined
We get equally distinguishable patterns when we run the victim on other cores, as well as on our Skylake machine (cf. Appendix~\ref{appendix:side-channel}).
\else
We get equally distinguishable patterns when we run the victim on other cores, as well as on our Skylake machine (see the extended version~\cite{extendedversion}).
\fi

\addtocounter{footnote}{-2} %
\stepcounter{footnote}
\footnotetext{
		When $bit=1$ an RSA victim's iteration lasts $T_\textit{E1+E2} = 11,230$ cycles, that allow the attacker to collect ${\footnotesize \sim} 51$ samples.
		When $bit=0$, it lasts $T_\textit{E1} = 5,690$ cycles and is followed by an interval of no contention (second call to $E1$); the sum of these intervals allows the attacker to collect ${\footnotesize \sim} 43$ samples. 
		To better compare the two traces, we cut both of them at $43$ samples.
}

\stepcounter{footnote}
\footnotetext{
    Iterations of the EdDSA victim ($T_\textit{E1+E2} = 35,120$ cycles and $T_\textit{E1} = 18,260$ cycles) take longer than the ones of the RSA victim. 
    Hence, the attacker is able to collect a larger number of samples. 
}

To train our classifier, we collect a set of 5000 traces, half of which with the victim operating on a 0 bit and the other half with it operating on a 1 bit.
We use the first 43 samples from each trace as input vectors, and the respective 0 or 1 bits as labels.
We then randomly split the set of vectors into 75\% training set and 25\% testing set, and train our classifier to distinguish between the two classes. %
Our classifier achieves an accuracy of 90\% with prefetchers on and 86\% with prefetchers off, %
demonstrating that a single trace of load latencies 
measured by the attacker during a victim's iteration 
can leak that iteration's secret key bit with high accuracy.

\paragraph{Results for EdDSA}
We target the EdDSA Curve25519 signing code of libgcrypt 1.6.3, which includes a secret-dependent code path in its elliptic curve point-scalar multiplication function \texttt{\_gcry\_mpi\_ec\_mul\_point}. %
In this function, the doubling phase represents \textit{E1}, executed unconditionally, and the addition phase represents \textit{E2}, executed conditionally only on 1-valued bits of the key (i.e., the scalar).

We report in Figure~\ref{fig:side-channel-eddsa-dreadnought} the results of leaking a bit using the same setup as in the RSA attack.
Both traces start with peaks corresponding to the first call to \textit{E1}.
However, only when the secret bit is 1 do we observe an additional peak on the right-hand side of the plot.
This additional peak corresponds to the call to \textit{E2}.
\ifx\shortversion\undefined
We get similar patterns with the victim on other cores, as well as on our Skylake machine (cf. Appendix~\ref{appendix:side-channel}).
\else
We get similar patterns with the victim on other cores, as well as on Skylake (see the extended version~\cite{extendedversion}).
\fi
We train our classifier like we did for the RSA attack, except that the individual vectors now contain 140 samples.
Our classifier achieves an accuracy of 94\% with prefetchers on and 90\% with prefetchers off.

\subsection{Keystroke Timing Attacks}

Our second side channel attack leaks the timing of keystrokes typed by a user.
That is, like prior work~\cite{ristenpart2009hey,jana2012memento, lipp2017practical, kurth2020netcat, zhang2009peeping, pessl2016drama,gruss2016flush+, vila2017loophole}, the goal of the attacker is to detect \textit{when} keystrokes occur and extract precise inter-keystroke timings.
This information is sensitive because it can be used to reconstruct typed words (e.g., passwords)~\cite{zhang2009peeping, song2001timing, kurth2020netcat}.
To our knowledge, this is the first time a contention-based microarchitectural channel (cf. Section~\ref{ss:side-channel-background}) has been used for keystroke timing attacks.

Our attack builds on the observation that handling a keystroke is an involved process that requires interaction of multiple layers of the hardware and software stack, including the southbridge, various CPU components, kernel drivers, character devices, shared libraries, and user space~processes~\cite{typingwithpleasure, pressingakeyraspberry, ptynicola, kerrisk2010linux, schwarz2018keydrown, monaco2018sok}.
Prior work has shown that terminal emulators alone incur keystroke processing latencies that are in the order of milliseconds~\cite{lwnterminalemulators, danluuterminallatency} (i.e., millions of cycles).
Moreover, handling even a single keystroke involves executing and accessing large amounts of code and data~\cite{schwarz2018keydrown}.
Thus, we hypothesize that, on an otherwise idle server, keystroke processing may cause detectable peaks in ring contention.

\begin{figure}[t]
	\centering
	\includegraphics[width=.92\columnwidth]{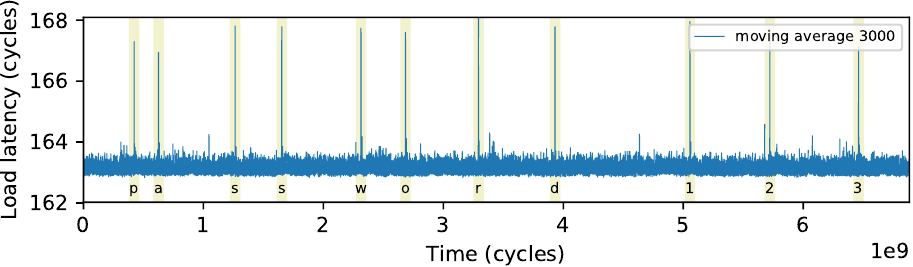}
	\caption{
		Load latency measured by the attacker when a user types \texttt{password123} on the terminal, with $R_c = 3$ and $R_s~=~2$ (on Coffee Lake). 
		Latency spikes occur reliably upon keystroke processing (yellow bars) and can be used to extract inter-keystroke timings.
		See Figure~\ref{fig:side-channel-keyboard-zoom-dreadnought} for a zoomed-in plot. %
	}
	\label{fig:side-channel-keyboard-dreadnought}
\end{figure}

\paragraph{Implementation}

To validate our hypothesis, we develop a simple console application that calls \texttt{getchar()} in a loop and records the time when each key press occurs, to serve as the ground truth (as in~\cite{zhang2009peeping}).
We consider two scenarios: i) typing on a local terminal (with physical keyboard input~\cite{typingwithpleasure, pressingakeyraspberry}), and ii) typing over an interactive SSH session (with remote input~\cite{kerrisk2010linux,barrett2005ssh}).

\paragraph{Results}

Figure~\ref{fig:side-channel-keyboard-dreadnought} shows a trace collected by the attacker on our Coffee Lake machine in the SSH typing scenario, after applying a moving average with a window of 3000 samples. 
We report a zoomed-in version of the trace for a single keystroke in Figure~\ref{fig:side-channel-keyboard-zoom-dreadnought}.
Our first observation is that when a keystroke occurs, we observe a very distinguishable pattern of ring contention.
Running our attack while typing the first 100 characters of the ``To be, or not to be'' soliloquy, we observed this pattern upon all keystroke events with zero false negatives and zero false positives.
Further, ring contention peaks were always observed well within 1 ms (\num{3e6} cycles)
of recording a keystroke, which is the precision required by the inference algorithms used by prior work to differentiate the keys pressed.
\ifx\shortversion\undefined
We got similar results when we typed keystrokes on a local terminal as well as on our Skylake machine (cf. Appendix~\ref{appendix:side-channel}).
\else
We got similar results when we typed keystrokes on a local terminal as well as on Skylake (see the extended version~\cite{extendedversion}).
\fi
Moreover, we tested our attack while running \texttt{stress -m N} in the background, which spawns \texttt{N} threads generating synthetic memory load on the system.
The results, reported in Appendix~\ref{appendix:side-channel}, show that the temporal patterns of ring contention on keystrokes 
were still easily distinguishable from background noise when $\texttt{N}\le2$.
However, as the load increased (with $\texttt{N} > 2$), keystrokes started to become harder to identify by simply using the moving average of Figure~\ref{fig:side-channel-keyboard-dreadnought}, and with $\texttt{N} > 4$, they started to become almost entirely indistinguishable from background noise.

We believe that the latency peaks we observe on keystrokes are caused by ring contention (and not, e.g., cache evictions or interrupts) for several reasons.
First, the latency differences caused by contention on keystrokes are in the same range~of the ones we measured in Section~\ref{s:reverse-engineering}.
Second, we observed~that, although keystroke processing creates contention on all slices, latency peaks are more pronounced when the attacker monitors ring segments that incur more contention (i.e., the tables with the most gray cells in Figures~\ref{fig:heatmap-no-miss} and~\ref{fig:heatmap-miss}).
For example, when $R_c = 0$ and $R_s = 7$ the peaks upon keystrokes are smaller than in most other configurations.
This is because in this configuration the attacker always has priority on both the request ring (there is no core upstream of $R_c$ whose request traffic can delay $R_c$'s one) and the data/acknowledge rings (there is no slice/SA 
upstream of $R_s$ whose data/acknowledge traffic can delay $R_s$'s one).
Hence, we believe the only contention~that occurs in this case is slice contention. %
Third, when we tried to repeat our experiments with the attacker timing L1 hits instead of LLC hits, we did not see latency peaks upon keystrokes.

\section{Discussion and Future Work}
\label{s:discussion}

\paragraph{Attack Requirements}
Our attack on cryptographic code (cf. Section~\ref{s:side-channel-crypto}) requires the victim's cache to be cleansed on context switches. 
On the one hand, this requirement limits the applicability of the attack, considering that cache cleansing is not currently done by major OSs.
On the other hand, however, cache cleansing is often recommended~\cite{sprabery2018scheduling,godfrey2013server, zhang2013duppel, varadarajan2014scheduler, ge2018survey, gullasch2011cache, braun2015robust, osvik2006cache, guanciale2016cache, ge2018no, ferraiuolo2018hyperflow,percival2005cache} as a defense against cache-based preemptive scheduling attacks, and may be deployed~in~the~future if temporal isolation starts getting added to OSs.
If so, defenders would be in a lose-lose situation: either they i)~do~not~cleanse and get attacked through preemptive scheduling attacks or ii) cleanse and get attacked through our attack. 
These results highlight that side channel mitigations still need more study.

Moreover, our attack POC assumes (like prior work~\cite{osvik2006cache,guanciale2016cache, bruinderink2016flush, evtyushkin2018branchscope,wang2019papp, evtyushkin2016understanding, aciiccmez2007yet, aciiccmez2008vulnerability}) the availability of preemptive scheduling techniques.
A real attack, however, would include an implementation of such techniques.
High-precision variants of these have been demonstrated for non-virtualized settings~in \cite{gullasch2011cache,roy2015design,ashokkumar2016highly,neve2006advances}, and shown to be practical in virtualized settings in~\cite{zhang2012cross}.\footnote{
	These techniques exploit the designs of the Linux/Xen CPU schedulers.
	The attacker spawns multiple threads, some of which run on the same CPU as the victim.
	The threads running on the victim's CPU sleep most of the time.
	However, at carefully chosen times the attacker wakes them up, causing the scheduler to interrupt the victim to run the attacker.
}
Preemptive scheduling is also practical against trusted execution environments such as Intel SGX~\cite{van2017sgx}.
Yet, future work is needed to assess the practicality of preemptive scheduling in more restricted environments such as browsers. 

\paragraph{Mitigations}
Intel classifies our attack as a ``traditional side channel'' because it takes advantage of architecturally committed operations~\cite{intel-timing-guidelines}.
The recommended line of defense against this class of attacks is to rely on software mitigations, and particularly on following constant-time programming principles.
The attacks we demonstrate on RSA and EdDSA rely on code that is not constant time; in principle, this mitigation should be effective in blocking them.
However, a more comprehensive understanding of hardware optimizations %
is needed before we can have truly constant-time code.
For~example, it was recently reported that Intel CPUs perform hardware store elimination between the private caches and ring interconnect~\cite{travis-down-store-elimination}.
This optimization may break constant-time programming by making ring contention a function of cache line \textit{contents}.

Further, additional mitigations are needed to block our covert channel and our keystroke timing attack.
Among hardware-only mitigations, designs based on spatial partitioning and statically-scheduled arbitration policies (e.g.,~\cite{wassel2013surfnoc}) could ensure that no ring contention can occur between processes from different security domains.
However, they would need additional mechanisms to mitigate slice contention.
Alternatively, less invasive hardware-software co-designs could be studied that allow ``trusted'' and ``untrusted'' code to only run on cores of different 
clusters (cf. Table~\ref{t:lane-agent}), and only access slices of different
clusters.
However, such approaches would require careful consideration to account for LLC misses, which may create traffic that crosses clusters. 

Finally, since our attacks rely on a receiver constantly missing in its private caches and performing loads from a target LLC slice, it may be possible to develop software-only anomaly detection techniques that use hardware performance counters to monitor bursts of load requests traveling to a single LLC slice.
However, these techniques would only be useful if they had small false positive rates.

\paragraph{Applicability to Other CPUs}
It should be possible to port our attacks on other CPUs using a ring interconnect.
\ifx\shortversion\undefined
For example, we were able to replicate our methodology on a server-class Xeon Broadwell CPU, finding that the distributed (``boxcar''-based) arbitration policy is the same that we observed on our client-class CPUs (more details are in Appendix~\ref{appendix:xeon-results}). %
\else
For example, we were able to replicate our methodology on a server-class Xeon Broadwell CPU, finding that the distributed (``boxcar''-based) arbitration policy is the same that we observed on our client-class CPUs (more details in the extended version~\cite{extendedversion}). %
\fi
An open question is whether our attacks can be generalized to CPUs that do not use a ring interconnect.
For example, recent server-class Intel CPUs utilize mesh interconnects~\cite{intel2017mesh}, which consist of a 2-dimensional array of half rings. %
Traffic on this grid-like structure is always routed vertically first and then horizontally.
More wires may make it harder for an attacker to contend with a victim.
At the same time, however, they may provide the attacker with more fine-grained visibility onto what segments a victim is using, but this topic merits further investigation.
Finally, AMD CPUs utilize other proprietary technologies known as Infinity Fabric/Architecture for their on-chip interconnect~\cite{amd-infinity-fabric, amd-infinity-architecture}.
Investigating the feasibility of our attack on these platforms requires future work. 
However, the techniques we use to build our contention model can be applied on these platforms too.

\section{Conclusion}
\label{s:conclusion}

In this paper, we introduced side channel attacks on the ring interconnect.
We reverse engineered the ring interconnect's protocols to reveal the conditions for two processes to incur ring contention.
We used these findings to build a covert channel with a capacity of over 4 Mbps, the largest to date for cross-core channels not relying on shared memory.
We also showed that the temporal trends of ring contention can be used to leak key bits from vulnerable EdDSA/RSA implementations as well as the timing of keystrokes typed by a user.
We have disclosed our results to Intel.

\section*{Acknowledgments}

This work was partially supported by NSF grants 1954521 and 1942888 as well as by an Intel ISRA center.
We thank our shepherd Yossi Oren and the anonymous reviewers for their valuable feedback.
We also thank Gang Wang for his valuable suggestions on early drafts of this paper, and Ben Gras for the helpful discussions on the first side channel POC.

\section*{Availability}

We have open sourced the code of all the experiments of this paper at \url{https://github.com/FPSG-UIUC/lotr}.

\bibliographystyle{plain}
{\footnotesize
\bibliography{main}}

\appendix
\section{Additional Data}
\label{appendix:extra-data}

\begin{figure}[t!]
\begin{lstlisting}
void **addr; /* circular pointer-chasing list of W_LLC addresses */
const int repetitions = 100000; /* number of samples wanted */
uint32_t samples[repetitions];	/* trace */
for (i = 0; i < repetitions; i++) {
	asm volatile(
		"lfence\n"
		"rdtsc\n"						/* eax = TSC */
		"mov %%eax, %%edi\n"			/* edi = eax */
		"mov (%1), %1\n"				/* addr = *addr; LOAD */
		"mov (%1), %1\n" 				/* addr = *addr; LOAD */
		"mov (%1), %1\n" 				/* addr = *addr; LOAD */
		"mov (%1), %1\n" 				/* addr = *addr; LOAD */
		"rdtscp\n"						/* eax = TSC */
		"sub %%edi, %%eax\n" 			/* eax = eax - edi */
		: "=a"(samples[i]), "+r"(addr)	/* samples[i] = eax */
		: : "rcx", "rdx", "edi", "memory"); 
}
\end{lstlisting}
	\setcaptiontype{lstlisting}
	\caption{Timed loads used to monitor the ring interconnect.}
	\label{lst:timing_code}
\end{figure}

\ifx\shortversion\undefined

\begin{figure}[t]
	
	\begin{subfigure}{\columnwidth}
		\centering
		\includegraphics[width=.92\columnwidth]{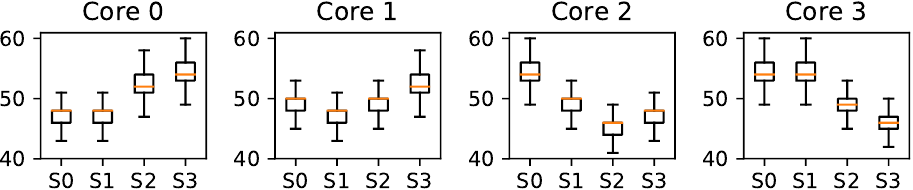}
		\caption{
			Using an even cache set $p=10$.
		}
		\label{fig:latency-skylake-even}
	\end{subfigure}
	
	\vspace{2pt}

	\begin{subfigure}{\columnwidth}
		\centering
		\includegraphics[width=.92\columnwidth]{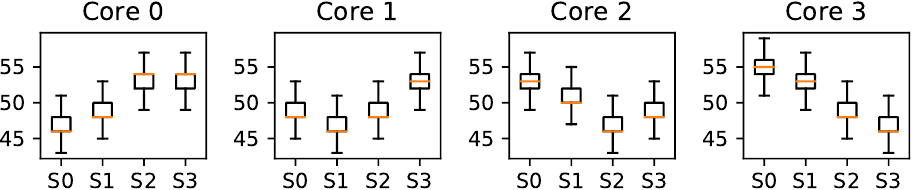}
		\caption{
			Using an odd cache set $p=5$.
		}
		\label{fig:latency-skylake-odd}
	\end{subfigure}

	\caption{
		Load latency (in cycles) for different combinations of core $c$, slice $s$ and set $p$ on our Skylake machine.
	}
	\label{fig:skylake-topology}
\end{figure}

\begin{figure}[t!]
	\centering
	\includegraphics[width=0.92\columnwidth]{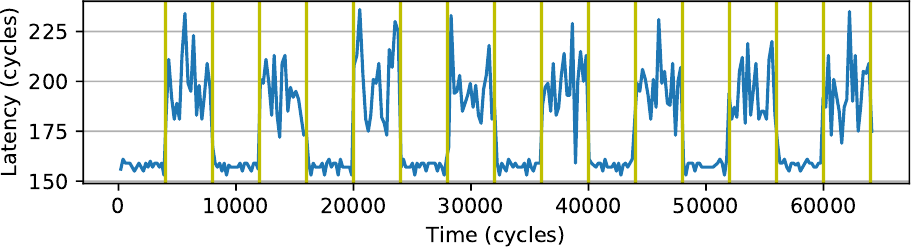}
	\caption{
		Load latency measured by our covert channel receiver when the sender transmits a sequence of zeros and ones on our Skylake machine, with $R_c = 2$, $R_s = 1$, $S_c = 1$, $S_s = 0$ and a transmission interval of 4,000 cycles.
	}
	\label{fig:covert-channel-skylake}
\end{figure}

\begin{figure}[t!]
	\centering
	\includegraphics[width=.92\columnwidth]{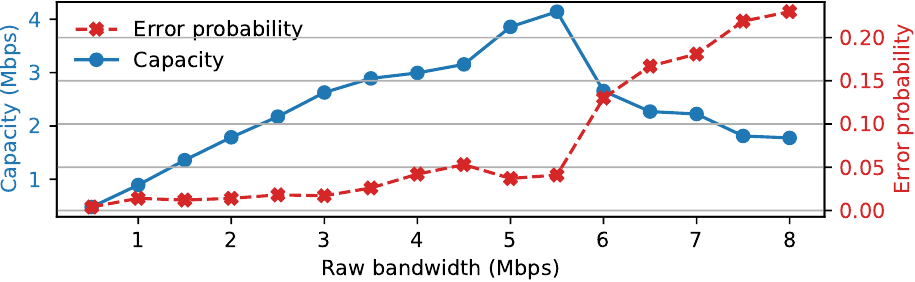}
	\caption{
		Covert channel performance on Skylake, using the same configuration of Figure~\ref{fig:covert-channel-skylake}. 
		We report raw bandwidth, error probability and channel capacity (as in Figure~\ref{fig:covert-channel-capacity-coffeelake}). %
	}
	\label{fig:covert-channel-capacity-skylake}
\end{figure}

\begin{figure*}[h!]
	\centering
	\includegraphics[width=.92\linewidth]{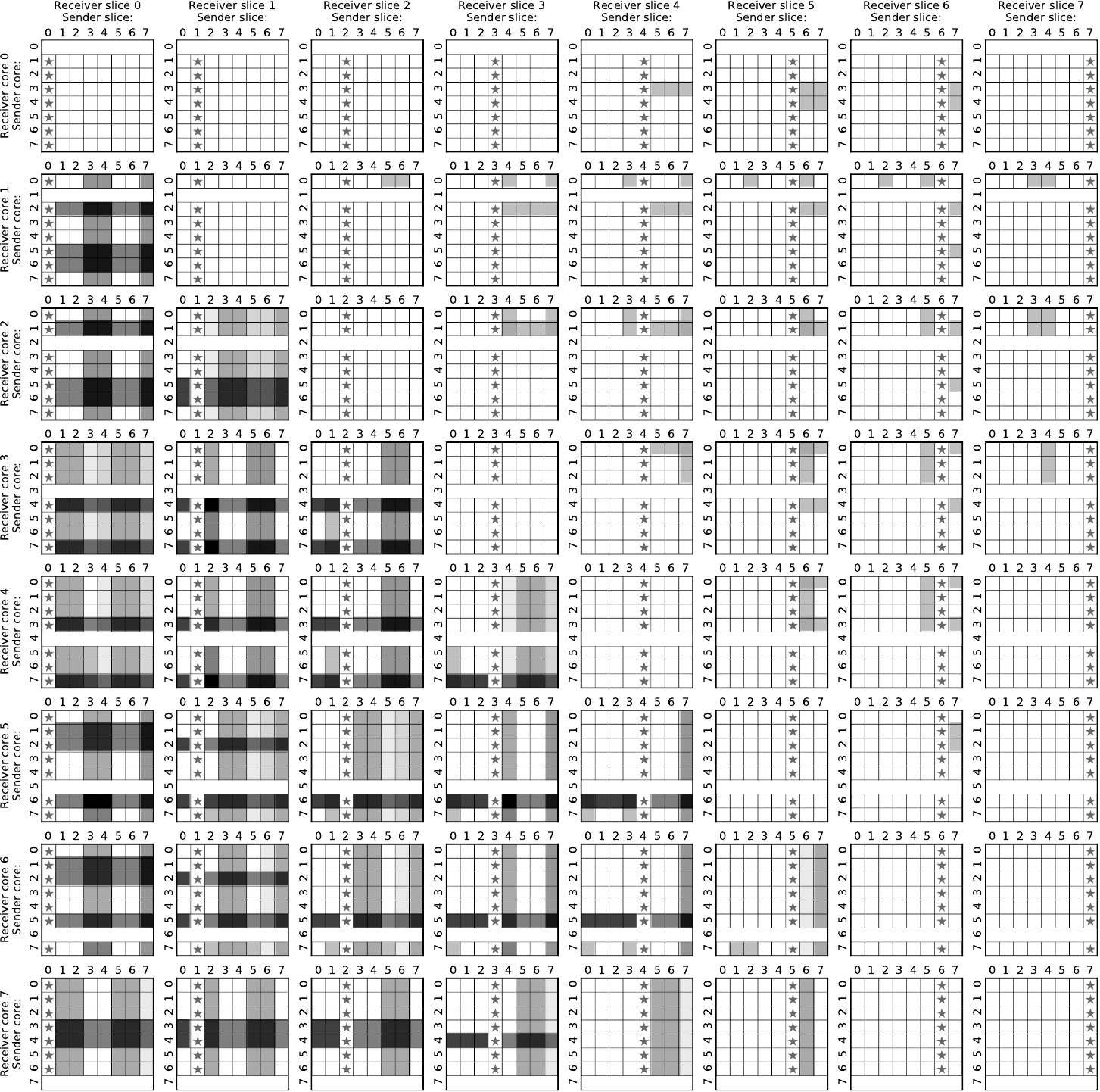}
	\caption{
		Ring interconnect contention heatmap when the receiver performs loads that hit in LLC, and the sender performs loads that miss in the LLC.
		Similar to Figure~\ref{fig:heatmap-no-miss}, the $y$ axes indicate the core where the sender and the receiver run, and the $x$ axes indicate the LLC slice from which they perform their loads.
		Cells with a star ($\bigstar$) indicate slice contention (when $R_s = S_s$), while \textbf{gray} cells indicate contention on the ring interconnect (with darker grays indicating larger amounts of contention).
	}
	\vspace{-5pt}
	\label{fig:heatmap-miss}
\end{figure*}

\fi

\ifx\shortversion\undefined

\else

We refer the reader to the extended version of this paper~\cite{extendedversion} for the full appendix, including more results from our Coffee Lake and Skylake CPUs, as well as preliminary results from our Xeon Broadwell CPU. 

\fi

\subsection{Reverse Engineering}
\label{appendix:reverse-engineering}

\ifx\shortversion\undefined

Figure~\ref{fig:skylake-topology} shows the load latency on our Skylake CPU for all core-slice combinations and with two different target cache sets.
These results, similarly to the ones for our Coffee Lake CPU that were shown in Figure~\ref{fig:latency-coffeelake}, confirm that the load latency grows when the loads have to travel a longer distance on the ring, which matches the logical ring topology of Figure~\ref{fig:ring-interconnect-diagram}.
We report an additional observation that we made while analyzing these results.
Only on our Skylake CPU, the load latency from a target slice sometimes also depends on the parity of the target cache set.
For example, the latency for core $c=0$ to perform a load from slice $s=0$ is smaller when the cache set $p$ is odd than when it is even.
In the slice-core combinations where this behavior occurs, we observe it systematically for any odd or even set.
This behavior does not affect our ring contention model.
We hypothesize that it may be due to some LLC cache banking \textit{within} individual LLC slices.

\fi

Figure~\ref{fig:heatmap-miss} reports the ring interconnect contention results when the sender misses in the LLC. 
Section~\ref{s:reverse-engineering} contains an in-depth analysis of these results (cf. Equation~\ref{eq:ring-contention-miss}).

\subsection{Covert Channel}
\label{appendix:covert-channel}

\ifx\shortversion\undefined

Figures~\ref{fig:covert-channel-skylake} and~\ref{fig:covert-channel-capacity-skylake} show the results of the covert channel on our Skylake machine. 
In particular, Figure~\ref{fig:covert-channel-skylake} reports a sample trace collected when the sender is transmitting a sequence of ones (peaks) and zeros (valleys) with a transmission interval of 4,000 cycles, showing that the two are visibly distinguishable. 
Observe that the configuration we picked this time ($R_c = 2$, $R_s = 1$, $S_c = 1$, $S_s = 0$) is an example of a configuration that sees contention only when the prefetchers are turned on, which would normally not be subject to contention if the prefetchers were off (cf. Equation~\ref{eq:ring-contention-no-miss}).
Figure~\ref{fig:covert-channel-capacity-skylake} reports the performance of our covert channel on Skylake, which reaches a maximum capacity of 4.14 Mbps (518 KBps) given a transmission interval of 727 cycles.

\else

Figure~\ref{fig:covert-channel-capacity-skylake} reports the performance of our covert channel on our Skylake machine, which reaches a maximum capacity of 4.14 Mbps (518 KBps) given a transmission interval of 727 cycles.
Observe that the configuration we picked this time ($R_c = 2$, $R_s = 1$, $S_c = 1$, $S_s = 0$) is an example of a configuration that sees contention only when the prefetchers are turned on, which would normally not be subject to contention if the prefetchers were off (cf. Equation~\ref{eq:ring-contention-no-miss}).

\fi

\subsection{Side Channels} 
\label{appendix:side-channel}

\ifx\shortversion\undefined

\begin{figure}[t]
	\begin{subfigure}{\columnwidth}
		\centering
		\includegraphics[width=0.87\columnwidth]{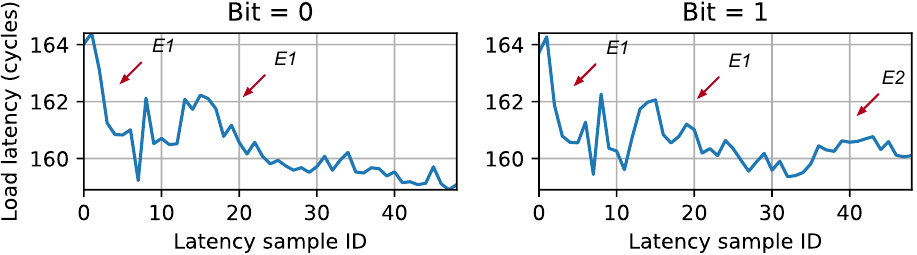}
		\caption{
			Results for the RSA victim.
			When the secret bit is 1, the attacker sees an additional contention peak between samples 30 and 45.
		}
		\label{fig:side-channel-rsa-skylake}
	\end{subfigure}
	
	\vspace{10pt}

	\begin{subfigure}{\columnwidth}
		\centering
		\includegraphics[width=0.87\columnwidth]{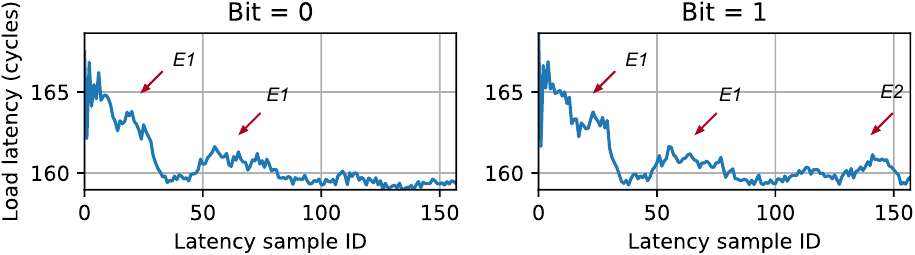}
		\caption{
			Results for the EdDSA victim.
			When the secret bit is 1, the attacker sees an additional peak after the 125-th sample.
		}
		\label{fig:side-channel-eddsa-skylake}
	\end{subfigure}

	\caption{
		Latencies measured by the attacker during a victim's iteration, with $R_c = 1$, $R_s = 0$, and $S_c = 2$ (on Skylake).
	}
	\label{fig:skylake-side-channels}
\end{figure}

\paragraph{Attack on Cryptographic Code}

Figures~\ref{fig:side-channel-rsa-skylake} and~\ref{fig:side-channel-eddsa-skylake} show traces (averaged over 100 runs) from our attack on RSA and EdDSA on our Skylake machine.
Both experiments are run with the attacker on core $R_c = 1$, timing loads from $R_s = 0$, and the victim on core $S_c = 2$.
The general trends resemble the ones of our Coffee Lake results (cf. Section~\ref{s:side-channel-crypto}).
For the attack against RSA, our classifier achieves an accuracy of 90\% with prefetchers on and 87\% with prefetchers off.
For the attack against EdDSA, it achieves an accuracy of 95\% with prefetchers on and 92\% with prefetchers off.

\else

\begin{figure}[t!]
	\centering
	\includegraphics[width=.94\columnwidth]{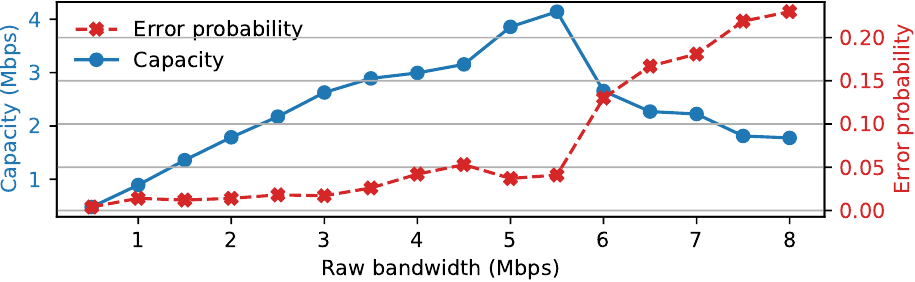}
	\caption{
		Covert channel performance on Skylake, with $R_c = 2$, $R_s = 1$, $S_c = 1$, $S_s = 0$. 
		We report raw bandwidth, error probability and channel capacity (as in Figure~\ref{fig:covert-channel-capacity-coffeelake}). %
	}
	\label{fig:covert-channel-capacity-skylake}
\end{figure}

\begin{figure}[t!]
	\vspace{4pt}
	\centering
	\begin{subfigure}{.475\columnwidth}
		\centering
		\includegraphics[width=\linewidth]{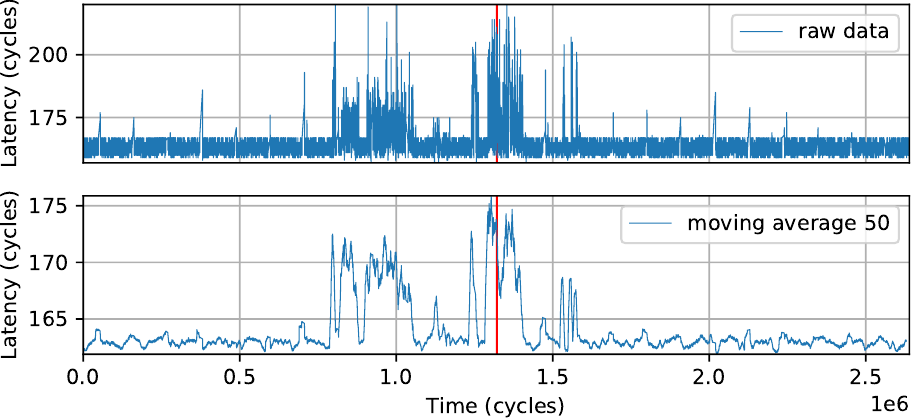}  
		\caption{No background noise.}
		\label{fig:side-channel-keyboard-zoom-dreadnought}
	\end{subfigure}
	\hspace{6pt}
	\begin{subfigure}{.475\columnwidth}
		\centering
		\includegraphics[width=\linewidth]{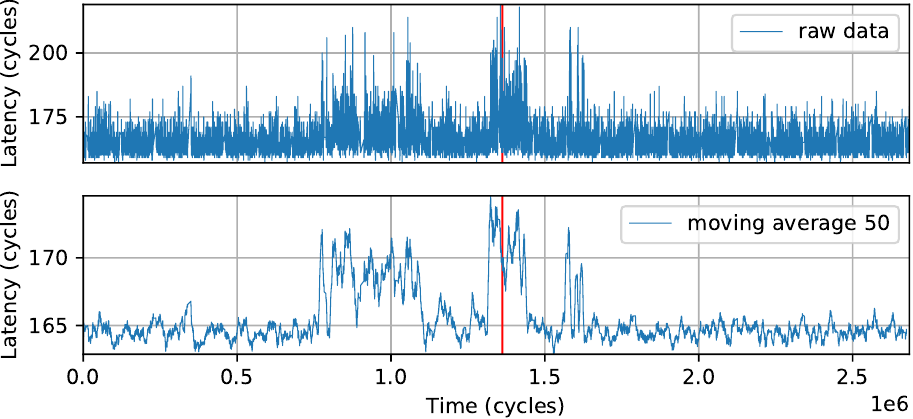}
		\caption{With \texttt{stress -m 1}.}
		\label{fig:side-channel-keyboard-zoom-dreadnought-stress}
	\end{subfigure}
	\\
	\vspace{8pt}
	\begin{subfigure}{.475\columnwidth}
		\centering
		\includegraphics[width=\linewidth]{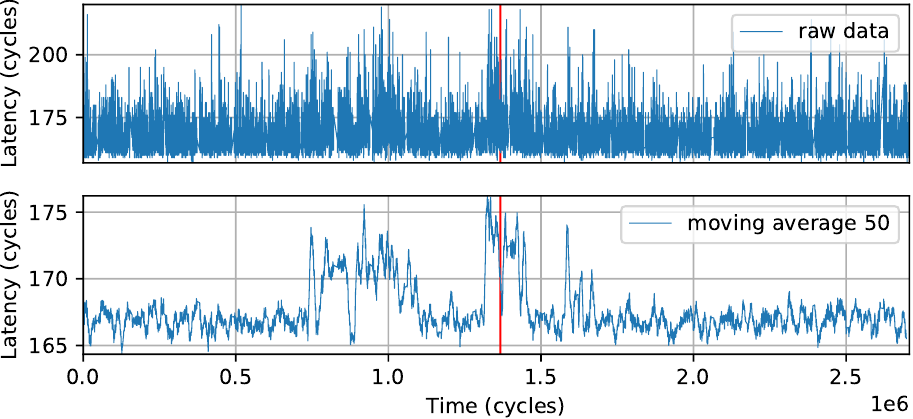}
		\caption{With \texttt{stress -m 2}.}
		\label{fig:side-channel-keyboard-zoom-dreadnought-stress2}
	\end{subfigure}
	\hspace{6pt}
	\begin{subfigure}{.475\columnwidth}
		\centering
		\includegraphics[width=\linewidth]{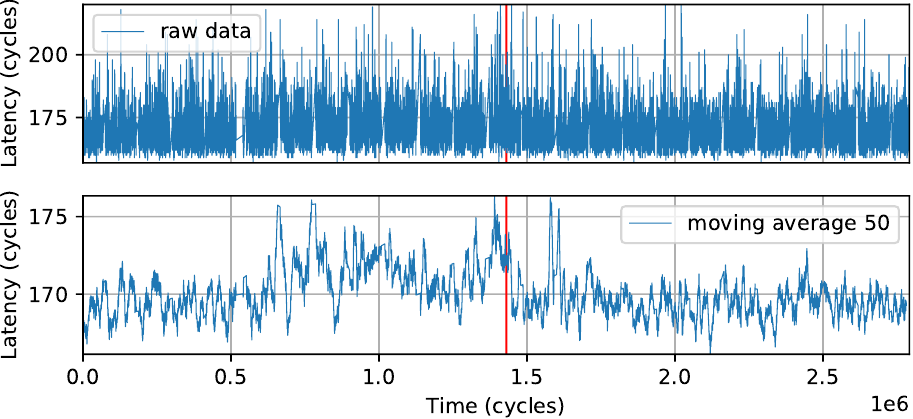}
		\caption{With \texttt{stress -m 4}.}
		\label{fig:side-channel-keyboard-zoom-dreadnought-stress4}
	\end{subfigure}
	\caption{
		Zoomed-in version of the load latencies measured by the attacker when a user types a single key on the terminal (on Coffee Lake), in the presence of different levels of background noise.
		The keystroke event was recorded in the middle (red line).
		We reliably observe the same ring contention pattern only upon keystroke events.
	}
	\label{fig:all-zooms-dreadnought}
\end{figure}

\begin{figure*}[h!]
	\centering
	\includegraphics[width=.98\linewidth]{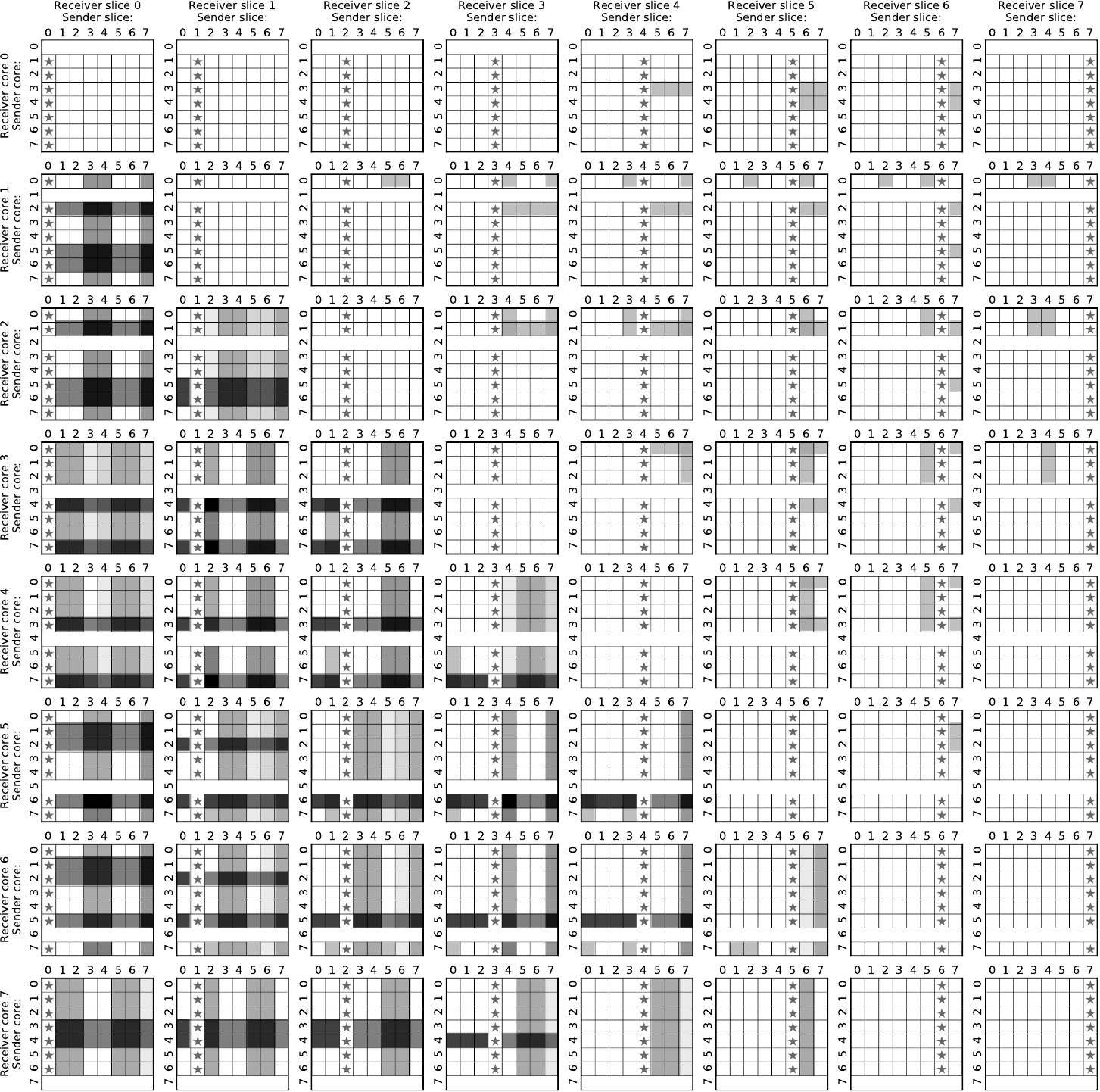}
	\caption{
		Ring interconnect contention heatmap when the receiver performs loads that hit in LLC, and the sender performs loads that miss in the LLC.
		Similar to Figure~\ref{fig:heatmap-no-miss}, the $y$ axes indicate the core where the sender and the receiver run, and the $x$ axes indicate the LLC slice from which they perform their loads.
		Cells with a star ($\bigstar$) indicate slice contention (when $R_s = S_s$), while \textbf{gray} cells indicate contention on the ring interconnect (with darker grays indicating larger amounts of contention).
	}
	\label{fig:heatmap-miss}
\end{figure*}

\fi

\ifx\shortversion\undefined

\paragraph{Cache Cleansing}

In our evaluation against cryptographic code, we implemented cache cleansing (to defend against preemptive scheduling attacks) by flushing the victim's cache lines with \texttt{clflush} during context switches.
Using \texttt{clflush} on every victim's cache line is the best we can do on current hardware (the only secure alternative would be to use \texttt{wbinvd} that invalidates all caches, as in~\cite{ge2018no}).
However, one may argue that with certain types of LLC partitioning (ensuring that victim and attacker do not share partitions even when time-sharing the same core), cleansing the LLC could be avoided.
In this section, we show that even if only the L1/L2 caches needed to be cleansed on context switches, our attack would still go through.
To this end, we repeat the attacks against RSA and EdDSA using an eviction-based L1/L2-only cleansing technique. 
To cleanse the L1d/L2, we iteratively access an eviction set with $W_{L2}$ addresses for each cache set of the victim's L2.
To cleanse the L1i, we execute a sequence of \texttt{jmp} instructions over a 32 KB region (as in~\cite{oleksenko2018varys,ge2019time}).\footnote{This L1/L2 flushing approach may be incomplete as it makes assumptions on the replacement policy. Intel does not currently provide an instruction to reliably remove a cache line from the L1d, L1i and L2 caches only.}

\begin{figure}[t]
	\begin{subfigure}{\columnwidth}
		\centering
		\includegraphics[width=.87\columnwidth]{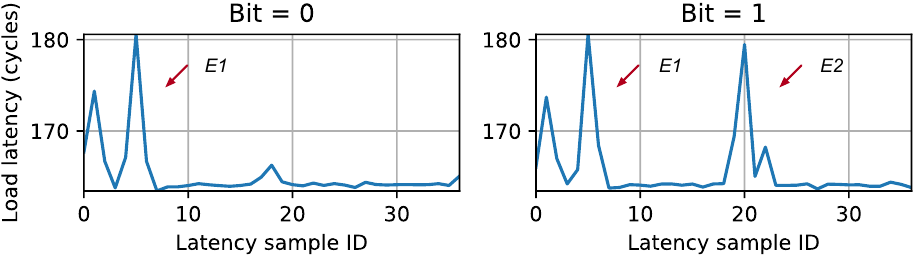}
		\caption{
			Results for the RSA victim.
		}
		\label{fig:side-channel-rsa-dreadnought-l1l2}
	\end{subfigure}
	
	\vspace{10pt}

	\begin{subfigure}{\columnwidth}
		\centering
		\includegraphics[width=.87\columnwidth]{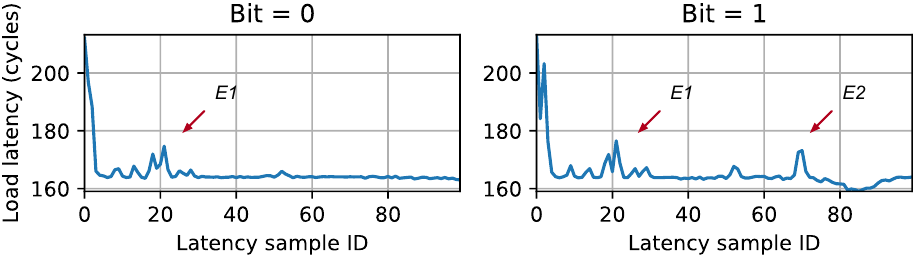}
		\caption{
			Results for the EdDSA victim.
		}
		\label{fig:side-channel-eddsa-dreadnought-l1l2}
	\end{subfigure}

	\caption{
		Latencies measured by the attacker during a victim's iteration, with $R_c = 5$, $R_s = 4$, and $S_c = 6$ (on Coffee Lake).
		This time, instead of flushing the victim's memory from the entire cache, we only evicted the L1/L2.
	}
	\label{fig:dreadnought-side-channels-l1l2}
\end{figure}

Figure~\ref{fig:side-channel-rsa-dreadnought-l1l2} and~\ref{fig:side-channel-eddsa-dreadnought-l1l2} show traces collected by the attacker to leak one key bit of each victim (averaged over 100 runs).
Because the victim only misses in the L1 and L2 but hits in the LLC, single iterations take less time than the ones we showed in Section~\ref{s:side-channel-crypto}, 
However, we can still see distinguishable ring contention patterns to recognize whether the secret bit was 0 or 1.
For example, when we train our classifier with $R_c = 5$, $R_s = 4$, and~$S_c = 6$ (on Coffee Lake, with prefetchers on), we achieve an accuracy of 93\% on RSA and 96\% on EdDSA.

\begin{figure*}[t!]
	\centering
	\begin{subfigure}{.23\textwidth}
		\centering
		\includegraphics[width=\linewidth]{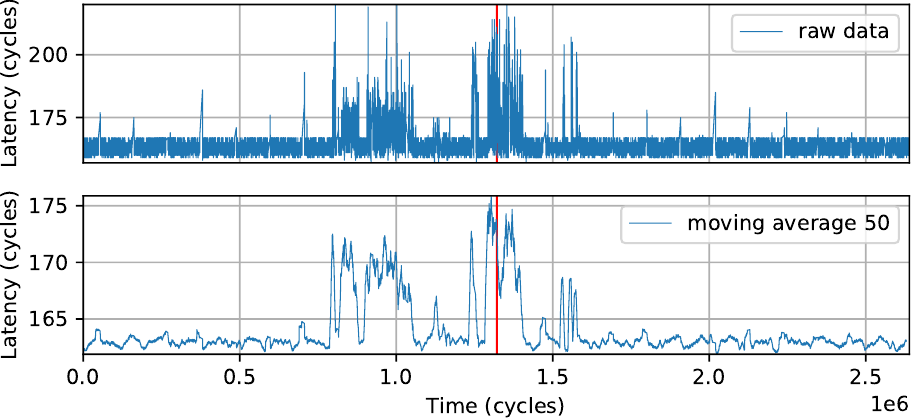}  
		\caption{No background noise.}
		\label{fig:side-channel-keyboard-zoom-dreadnought}
	\end{subfigure}
	\hspace{8pt}
	\begin{subfigure}{.23\textwidth}
		\centering
		\includegraphics[width=\linewidth]{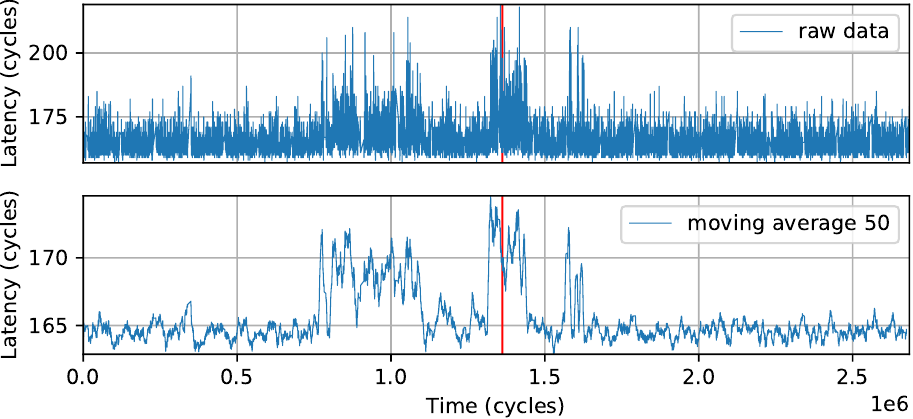}
		\caption{With \texttt{stress -m 1}.}
		\label{fig:side-channel-keyboard-zoom-dreadnought-stress}
	\end{subfigure}
	\hspace{8pt}
	\begin{subfigure}{.23\textwidth}
		\centering
		\includegraphics[width=\linewidth]{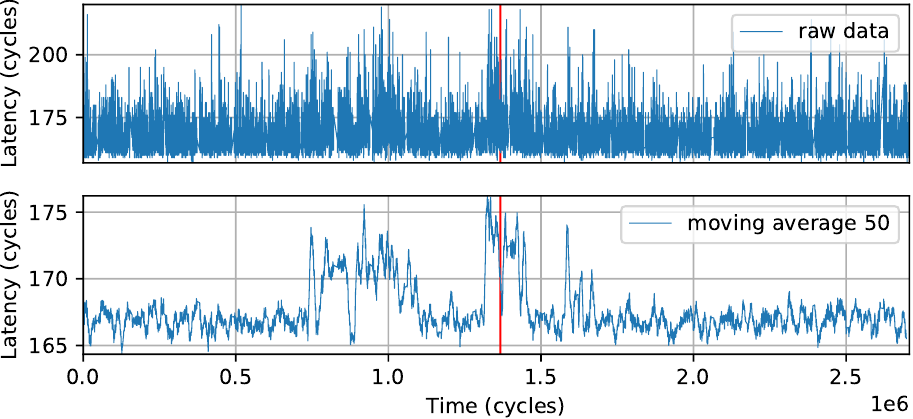}
		\caption{With \texttt{stress -m 2}.}
		\label{fig:side-channel-keyboard-zoom-dreadnought-stress2}
	\end{subfigure}
	\hspace{8pt}
	\begin{subfigure}{.23\textwidth}
		\centering
		\includegraphics[width=\linewidth]{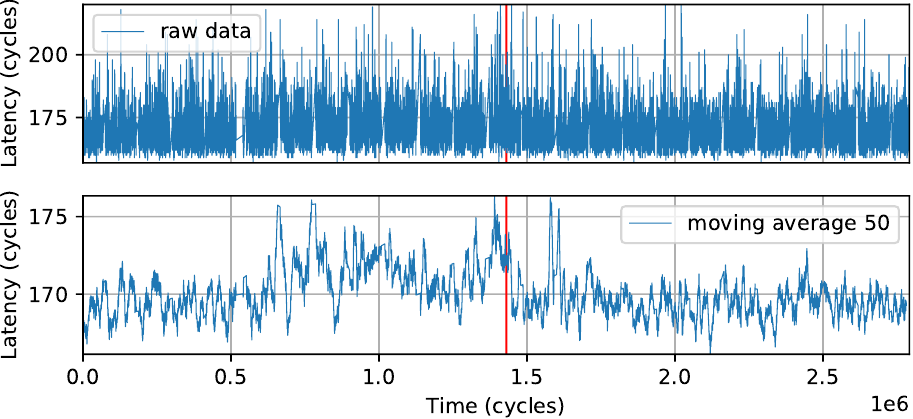}
		\caption{With \texttt{stress -m 4}.}
		\label{fig:side-channel-keyboard-zoom-dreadnought-stress4}
	\end{subfigure}
	\caption{
		Zoomed-in version of the load latencies measured by the attacker when a user types a single key on the terminal (on Coffee Lake), in the presence of different levels of background noise.
		The keystroke event was recorded in the middle (red line).
		We reliably observe the same ring contention pattern only upon keystroke events.
	}
	\label{fig:all-zooms-dreadnought}
\end{figure*}

\begin{figure}[t]
	\centering
	\includegraphics[width=.93\columnwidth]{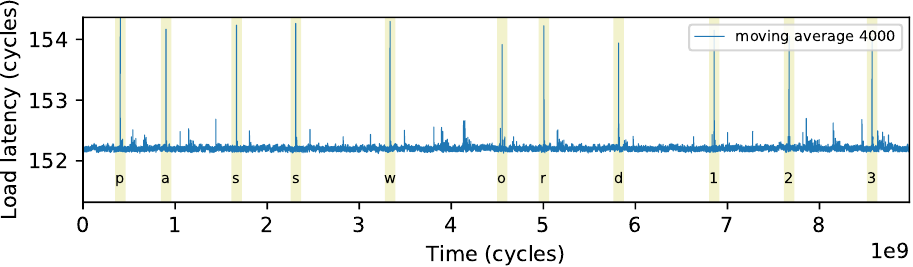}
	\caption{
		Latencies when the user types \texttt{password123} on Skylake (with $R_c = 1$ and $R_s = 0$).
		These results were collected in the SSH setting and in absence of background noise. 
	}
	\label{fig:side-channel-keyboard-skylake}
\end{figure}

\begin{figure}[t!]
	\centering
	\includegraphics[width=.93\columnwidth]{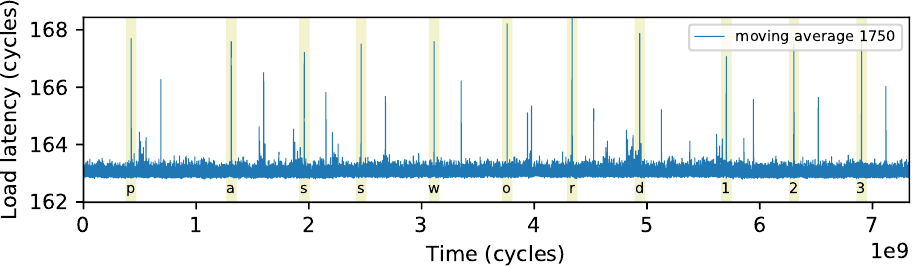}
	\caption{
		Latencies when the user types \texttt{password123} on Coffee Lake, in the local typing setting (with physical keyboard input) and in absence of background noise. 
	}
	\label{fig:side-channel-keyboard-dreadnought-local}
\end{figure}

\begin{figure}[t]
	\centering
	\vspace{2pt}
	\begin{subfigure}{.47\columnwidth}
		\centering
		\includegraphics[width=\columnwidth]{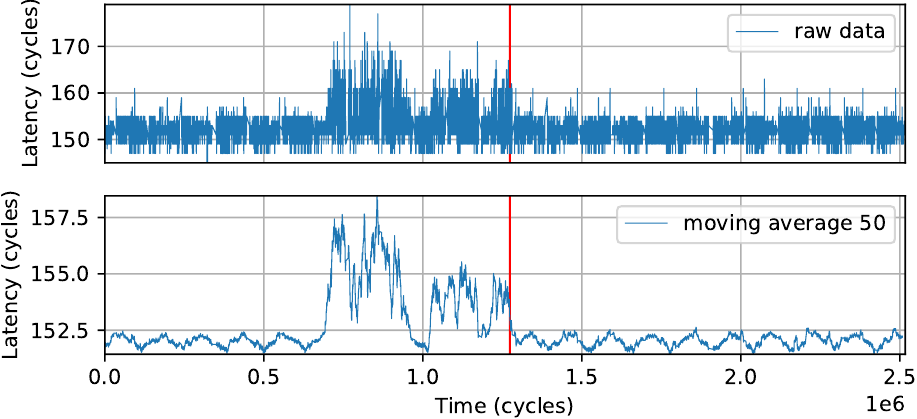}
		\caption{Skylake (SSH).}
		\label{fig:side-channel-keyboard-zoom-skylake}
	\end{subfigure}
	\hspace{8pt}
	\begin{subfigure}{.47\columnwidth}
		\centering
		\includegraphics[width=\columnwidth]{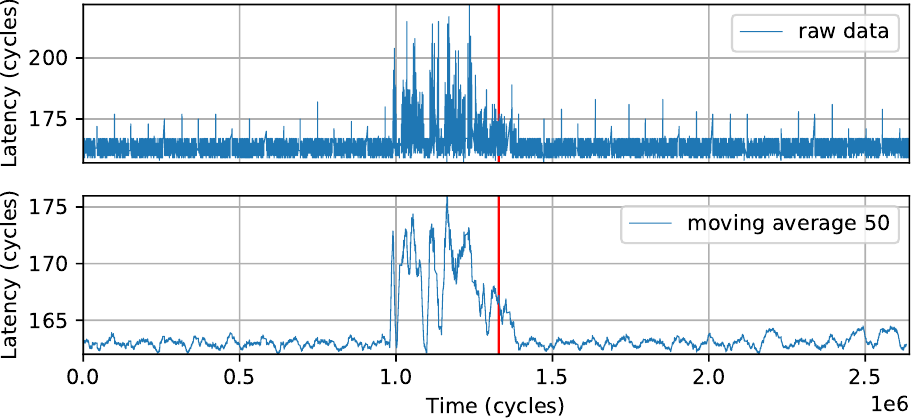}
		\caption{Coffee Lake (local).}
		\label{fig:side-channel-keyboard-dreadnought-local-zoom}
	\end{subfigure}
	\caption{
		Zoomed-in versions of Figures~\ref{fig:side-channel-keyboard-skylake} and~\ref{fig:side-channel-keyboard-dreadnought-local}.
	}
	\label{fig:other-zooms}
\end{figure}

\fi

\paragraph{Keystroke Timing Attack} 

Figure~\ref{fig:side-channel-keyboard-zoom-dreadnought} is a version of Figure~\ref{fig:side-channel-keyboard-dreadnought} zoomed-in to show the precise contention pattern that occurs upon keystroke events.
Figure~\ref{fig:side-channel-keyboard-zoom-dreadnought-stress} shows a trace collected while running \texttt{stress -m N}, with $\texttt{N}=1$ in the background.
Observe that keystroke events are still easily distinguishable from the background noise.
We get similar results when we run $\texttt{N}=2$ (Figure~\ref{fig:side-channel-keyboard-zoom-dreadnought-stress2}).
However, as the noise grows, with $\texttt{N} > 2$, we observe that keystroke events become harder to identify using a simple moving average (Figure~\ref{fig:side-channel-keyboard-zoom-dreadnought-stress4}).
Further, with $\texttt{N} > 4$, we observe that the keystroke events become almost entirely indistinguishable from noise.

\ifx\shortversion\undefined

Figures~\ref{fig:side-channel-keyboard-skylake} and~\ref{fig:side-channel-keyboard-zoom-skylake} show the results of the keystroke timing attack on our Skylake machine.
Similarly to Coffee Lake, we observe distinguishable contention patterns in the absence of noise.
We also repeat the experiments with background noise like above.
We observe that with $\texttt{N} \le 2$, keystrokes are still easily distinguishable from noise, but with $\texttt{N} = 3$ they get harder to identify using a simple moving average.

Figures~\ref{fig:side-channel-keyboard-dreadnought-local} and~\ref{fig:side-channel-keyboard-dreadnought-local-zoom} show the results of the attack when keystrokes are typed on a local terminal (with physical keyboard input\footnote{We tested the local typing scenario both on a \texttt{tty} and on a \texttt{pty}. We report results for the \texttt{tty} (Linux console) case, which is the harder case to attack (less code is executed to handle a keystroke on a \texttt{tty} than on a \texttt{pty}~\cite{ttydemystified}). However, we were able to successfully run the attack in the \texttt{pty} case too.}) on Coffee Lake.
We observe that keystrokes are still easily distinguishable, albeit the precise contention patterns look different than the ones of the SSH typing scenario.
One additional observation is that each keystroke-related peak in Figure~\ref{fig:side-channel-keyboard-dreadnought-local} is followed by a second, smaller peak: we hypothesize that this second peak may be due to the processing of the key release event.\footnote{In contrast, key release events are not transmitted over SSH.}
We also run the attack with background noise and observe results analogous to the ones above.
We get similar results on our Skylake machine too.

\fi

\ifx\shortversion\undefined

\begin{figure}[t!]
	\centering
	\vspace{5pt}
	\includegraphics[width=0.95\columnwidth]{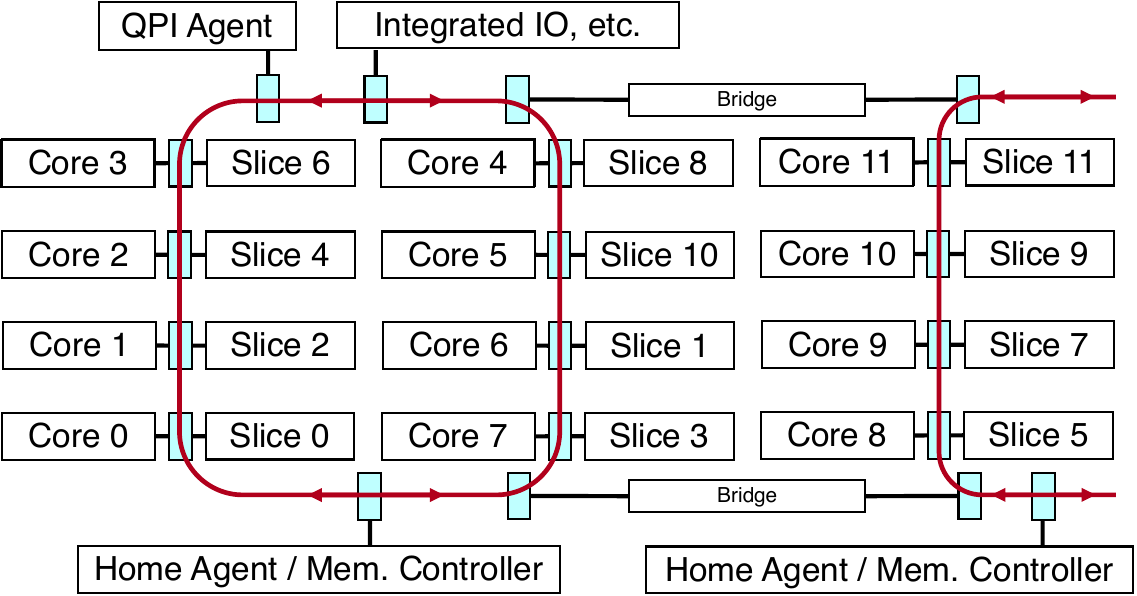}
	\caption{
		Logical block diagram of the ring interconnect on our 12-core server-class CPU.
		This diagram (except for the positioning of the slice indices, that come from our experimental results), comes from official Intel sources~\cite{intel-optimization-reference-manual, intel-xeon-e5-ring, intel-xeon-e5-uncore-monitoring}.
		We use the same color scheme as in Figure~\ref{fig:ring-interconnect-diagram}.
	}
	\label{fig:ring-interconnect-broadwell-diagram}
\end{figure}

\subsection{Results from a Server-class CPU} 
\label{appendix:xeon-results}

We now discuss preliminary results on a 12-core Intel Xeon E5-2687W v4 (Broadwell) CPU at 3.00 GHz.
Figure~\ref{fig:ring-interconnect-broadwell-diagram} shows the ring topology we inferred from our analysis on such server-class CPU.
This topology is different than the linear one of client-class CPUs (cf. Figure~\ref{fig:ring-interconnect-diagram}) in that it may include two rings (connected by bridges) instead of one.
In particular, the CPU we tested featured one and a half rings, matching the high-level diagram reported in official sources~\cite{intel-optimization-reference-manual, intel-xeon-e5-ring, intel-xeon-e5-uncore-monitoring} for a ``Medium Core Count (MCC)'' configuration.
Despite the topological differences, however, the distributed (``boxcar'' based) arbitration policy is the same.
As a result, we were still able to launch our attacks.
For example, with $R_c = 10$, $R_s = 7$, $S_c = 11$ and $S_s = 5$, we were able to successfully run our covert channel. 
Similarly, with $R_c = 10$, $R_s = 7$ and $S_c = 11$, we were able to replicate our side channel attack on cryptographic code, and with $R_c = 10$, $R_s = 7$ we were able to replicate our keystroke timing attack.
We found other sender/receiver configurations that lead to successful attacks too.
However, we leave it to future work to exhaustively reverse engineer the ring agent clustering rules and derive the equivalents of Equations~\ref{eq:ring-contention-no-miss} and~\ref{eq:ring-contention-miss} for server-class CPUs.
Finally, we made an additional observation on our Xeon CPU: under certain receiver configurations, the collected traces (including the effects of contention) sometimes scale up or down by constant offsets.
We hypothesize that this behavior may be due to Uncore Frequency Scaling (UFS), which was added to Xeon CPUs with the Haswell microarchitecture~\cite{ufsfrequencyscaling}.
However, this scaling is easy to account for due to its constant offset nature (when it kicks in, it looks like a step function).

\fi

\end{document}